\documentclass[aps,pra,preprint,groupedaddress,showpacs]{revtex4}
\usepackage{graphicx}% Include figure files
\usepackage{dcolumn}% Align table columns on decimal point
\usepackage{bm}% bold math
\begin{document}

\title{A Weyl function approach to matter-wave coherence and Talbot-Lau effects}
\affiliation{Department of Physics, Harvard University, Cambridge,
MA, 02138}
\author{Saijun Wu, Pierre S. Striehl, Mara G. Prentiss}

\affiliation{Department of Physics, Harvard University, Cambridge,
MA, 02138}

\affiliation {School of Engineering and Applied Science, Harvard
University, Cambridge, MA, 02138}

\date{\today}

\begin{abstract}
Weyl functions conveniently describe the evolution of wave
coherences in periodic or quadratic potentials. In this work we use
Weyl functions to study the ``Talbot-Lau effect'' in a time-domain
matter-wave interferometer. A ``displacement diagram'' is introduced
to analyze and calculate the matter-wave interference for an atomic
cloud in a quadratic potential that interacts with a sequence of
short optical standing wave pulses producing an atomic grating echo.
Unlike previous treatments, this new approach allows the atomic
ensemble to have an arbitrary initial phase-space distribution, and
the standing wave grating vectors to span three dimensions. Several
examples are discussed to illustrate the convenience of the
diagrammatic technique including the following: a two-dimensional
Talbot-Lau effect, the shift in the echo time and the recoil phase
for the interferometer perturbed by a quadratic potential; and the
realization of a time-domain ``Lau effect'' using a pulsed harmonic
potential. The diagrammatic technique is applicable to diffraction
gratings with arbitrary grating transmission functions. We conclude
the paper with a general discussion on the Weyl function
representations of matter-wave coherence, and relate the
conservation of matter-wave coherence with the conservation of
purity $\varsigma={\rm Tr}(\hat \rho^2)$ that distinguishes
decoherence effects from dephasing effects.
\end{abstract}


\maketitle

\addcontentsline{toc}{section}{} \tableofcontents

\section{Introduction}
%Wigner distribution function is convenient for establishing
%quantum-classical correspondence and for visualizing quantum
%distributions in phase space. The Fourier transform of the Wigner
%function is the Weyl function, that describes long-range orders of
%Wigner distribution in phase space. Wigner function method has
%received much attentions. In comparison Thought it was pointed out
%Weyl function as an important tool to study In this work we show
%that Weyl function conveniently describes the matter-wave dynamics
%in periodic or quadratic potentials.

In quantum phase space methods, operators in a Hilbert space are
mapped to functions in a ``mock'' phase
space~\cite{mockphasespace84}. One example is the Wigner
representation which is particularly useful for establishing
quantum-classical correspondence. In the Wigner representation, the
Wigner function represents a density matrix operator $\hat \rho$
with a quasi-probability distribution in classical phase space. The
Wigner functions are conveniently applied in semiclassical
methods~\cite{wignerheller76, berrywigner77}, to visualize or
characterize quantum states, and experimentally allows one to
recover information of quantum systems through tomographic
measurements of their phase-space distributions
~\cite{wignertomo93}.

In the related Weyl function representation~\cite{mockphasespace84,
dispoperators}, the Hilbert space operators are mapped to functions
in a ``reciprocal phase space'' where each function is simply a
Fourier transform of those in the Wigner representation. In
particular, Weyl function $W({\bf q},{\bf x})$ is the Fourier
transform of the Wigner function $w({\bf r},{\bf p})$ over the phase
space coordinates~$({\bf r},{\bf p})$~\cite{analyticrepresentations,
weylChountasis}. The Weyl function can also be expressed as:
\begin{equation}
\begin{array}{c}
W({\bf q},{\bf x})={\bf Tr}(\hat \rho \hat D({\bf q},{\bf x})),
\\ \hat D({\bf q},{\bf x})  = e^{i ({\bf q}\cdot \hat {\bf r} - {\bf
x}\cdot \hat {\bf p})}.
\end{array} \label{weyldefine}
\end{equation}
Here $\hat {\bf r}$ and $\hat {\bf p}$ are the position and momentum
operators, and $\hat D({\bf q},{\bf x})$ is the displacement
operator~\cite{dispoperators}, that shifts the density matrix by
$\hbar {\bf q}$ in momentum and ${\bf x}$ in position. Thus the Weyl
function $W({\bf q},{\bf x})$ is a phase-space correlation function
that measures the coherence of matter-waves. Though both Weyl and
Wigner functions completely specify any density matrix operators,
application of Weyl function, or the Fourier transform of the Wigner
function, has been much less discussed before~\cite{weylChountasis}.

In this work, we demonstrate the power of the Weyl function
technique by applying it to an important class of wave interference
phenomena, the Talbot-Lau effects associated with matter-waves in
quadratic or periodic potentials. Various examples of Talbot-Lau
type interference, including those closely related to recent
experiments~\cite{longcoherence,movingguide,mythesis}, are
discussed. To our knowledge, this work gives the first vectorial and
diagrammatic formula of Talbot-Lau effects. Apart from the expected
application in matter-wave optics, the formula may also provide a
convenient framework for the design of various lens-less imagine
systems~\cite{fish06}.

\subsection{Talbot-Lau effects}

Periodic light wavefronts can repeatedly reconstruct themselves
during free propagation. The phenomena is known as the Talbot effect
that was discovered by Henry F. Talbot~\cite{TalbotOrigin} in 1836.
A demonstration of the Talbot effect usually starts with a
collimated monochromatic light with wavelength $\lambda$. The light
passes through a transmission grating with a grating constant $d$ to
generate a periodic wavefront. Full or partial re-constructions of
the transmission wavefront at integer or fractional multiples of the
Talbot distance $z_T=d^2/\lambda$ are observed downstream. Many
intricate and elegant features of the Talbot phenomena have been
discussed in the literature~\cite{berrytalbot}. Instead of using a
collimated or spatially coherent input, in 1948 Ernst
Lau~\cite{LauOrigin} found that spatially incoherent light,
propagating through two gratings with same grating constant $d$
separated by a distance $z_{12}$, can generate an interference
pattern at the focal plane of a lens (Fraunhofer diffraction) placed
downstream from the second grating. The contrast of the interference
is maximized if $z_{12}$ takes integer multiples of $z_T/2$. More
general than Fraunhofer diffraction, in the Lau effect the
interference pattern appears at a distance $z_{23}=p/q z_{12}$
downstream from the second grating where $p,q$ are two integers.
Lau's original discovery corresponds to $p/q=N \rightarrow
\infty$~\cite{LauOrigin, laueffect79}.

Instead of using light waves, both the Talbot~\cite{PritchardTL,
BECTalbot99} and the Lau effects~\cite{potassiumTL} have been
demonstrated using matter-waves and in particular using neutral
atoms. Interferometry with multiple gratings using the Lau effect is
referred to as Talbot-Lau or Talbot-vonLau Interferometry (TLI). The
Lau effect is of particular interest for observing matter-wave
interference in the circumstances where a bright and collimated
matter-wave input is not available. Talbot-Lau interferometers are
used to demonstrate wave properties of atoms and
molecules~\cite{AIBerman, PritchardTL, potassiumTL, C70Inf,
BioTL03}, to study decoherence ~\cite{C70decohere03,
C70thermaldeco05} and to measure the polarizability of large
molecules~\cite{C60metro}. Talbot-Lau interferometers have also been
explored to sense inertial forces with
matter-waves~\cite{HanschEquivalence05, TLCahn97, GradientefTL06,
movingguide,mythesis}.

Based on Fresnel diffraction, matter-wave Talbot and Lau effects can
also happen in the time domain instead of in the spatial domain, due
to the equivalence of Sch{\"o}dinger's equation and the paraxial
wave equations. Matter-wave Talbot~\cite{BECTalbot99} and
Lau~\cite{TLCahn97} effects in the time domain have been observed
with atoms cooled down to a thermal velocity of centimeters per
second or lower. In the time domain, spatially separated gratings
are replaced by successive pulses of optical standing waves acting
as either phase or amplitude gratings for the matter-waves.
Time-domain interferometry using optical standing wave pulses with
laser-cooled atoms were pioneered by the authors of
ref.~\cite{TLCahn97}. In that experiment, off-resonant standing wave
pulses act as thin phase gratings to an atomic cloud. After each
standing wave pulse the atoms bunch to the potential minima of the
standing wave to form a density grating. The density grating washes
out due to velocity spreading of atoms. However, with two successive
standing wave pulses separated by time $T_{12}$ applied to atoms,
the atomic density grating revives around time $t=T_{12}+T_{23}$ if
$T_{23}=T_{12}$ and more generally if $T_{23}=N T_{12}$, where $N$
is an integer. The contrast of the revived density grating is a
periodic function of $T_{12}$, similar to the spatial Lau effect.
The revived atomic density grating is probed using a
Bragg-scattering of a probe light to induce a grating
echo~\cite{gratingechoOrigin}, and the interferometry based on this
setup is referred as a time-domain grating echo
interferometer~\cite{GradientefTL06}. Experiments and the theory of
the interferometry were further developed in ref.~\cite{Periodic02}
that generalizes the technique to include multiple standing wave
pulses. Recently a direct spatial image of sub-wavelength spatial
interference pattern due to the Lau effect after two standing wave
pulses (act as either amplitude or phase gratings) is achieved using
an on-resonant standing wave mask~\cite{TLturlapov05,Alexey06}.

Developed theories of matter-wave Talbot-Lau effect (TLE) usually
simplify the analysis by making particular assumptions on the
coherence properties of input matter-waves. Closely related to the
spatial (beam) Talbot-Lau setup with amplitude transmission
gratings, in the first approach ~\cite{KimbleTLTheory95,
ZeilingerTLTheory02, ClauserInftheory92} the atom flux through each
opening of the first grating are treated essentially as individual
point sources, from which the Fresnel diffraction on the image plane
is calculated. An incoherent sum of the contribution from all the
openings at the $1^{st}$ grating gives the final interference
pattern. Instead of incoherent summation of point sources, the
second approach~\cite{AIBerman, Periodic02, Resonantsplitter99}
assumes the input atoms as an incoherent sum of plane waves.
However, a general theory of Talbot-Lau interferometry applicable to
matter-waves with arbitrary initial spatial coherence hasn't been
presented in the literature.

The approach to TLE in this work is motivated by an obvious link
between the Weyl function [Eq.~(\ref{weyldefine})] and the
time-domain matter-wave TLE due to a sequence of standing wave
pulses. In the following we consider a single atom picture with
atomic motion characterized by a density matrix operator $\hat
\rho$. TLE corresponds to a revival of atomic density grating at a
certain time t. Since the revived density grating is spatially
periodic, we shall consider its particular $\bf Q$ component given
by:
\begin{equation}
\rho_{\bf Q}={\bf Tr}(\hat \rho(t) e^{i {\bf Q}\cdot \hat {\bf
r}})=W({\bf Q},0,t),\label{rhoQdefine}
\end{equation}
where $W({\bf q},{\bf x},t)$ is the Weyl function defined in
Eq.~(\ref{weyldefine}).

To express Eq.~(\ref{rhoQdefine}) in terms of an arbitrary initial
condition described by $\hat \rho(0)$, one can propagate $\hat
\rho(0)$ forward in time by considering the interaction of atoms
with a sequence of standing wave pulses. Alternatively, we shall
define the time-dependent displacement operator $\hat D({\bf q},{\bf
x},t) = e^{i ({\bf q}\cdot \hat {\bf r}(t) - {\bf x}\cdot \hat {\bf
p}(t))}$ in the Heisenberg picture and propagate the displacement
operator $\hat D({\bf Q},0,t)$ backward in time. Here we take the
latter approach which is advantageous for two reasons: 1) It applies
to general initial conditions. 2) As we will see, the equation of
motion for $\hat D({\bf q},{\bf x},t)$ due to pulsed periodic
potentials, during free-propagations and due to a quadratic
potential can be integrated easily. Since our analysis is generally
applied to different types of interferometers using TLE, we shall
generally consider an interferometry configuration using TLE as a
Talbot-Lau Interferometer, that will be referred to with the
abbreviation TLI in this work.

\subsection{Outline}
In what follows the paper is organized into three sections. In the
first section we introduce a diagrammatic technique to calculate the
grating echo of an N-pulse TLI in free-space. In the second part we
discuss the application of the theory to matter-waves in a quadratic
potential. In the discussion section we extend the formula to
include arbitrary grating transmission functions, discuss the Weyl
function representation of matter-wave coherence, and compare the
approach here with other diagrammatic techniques
~\cite{billiardball82}.

\section {Grating echo due to TLE in free space \label{firstsection}}

\subsection {The model \label{sectionmodel1}}
\begin{figure}
\centering
\includegraphics [width=5in,angle=0] {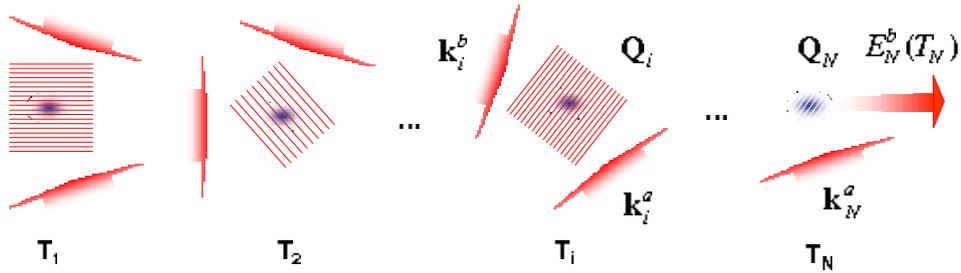}\\
\caption{Schematic of an N-pulse grating echo Talbot-Lau
interferometer}\label{gratingechoTLIfigure}
\end{figure}

In this section we consider TLE due to the simplest possible
interaction between 2-level atoms and a sequence of off-resonant
standing wave light pulses. The scenario is summarized in
Fig.~\ref{gratingechoTLIfigure}: an atomic sample localized at the
waists of multiple standing wave fields is subjected to a sequence
of N-1 standing wave pulses. At time $T_N$ a grating echo is induced
with the amplitude proportional to a particular spatial Fourier
component of the revived atomic density grating. The assumptions in
the model are summarized below. For notational convenience, we set
$\hbar = m = 1$.
\begin{enumerate}
\item[A1.]{\indent
2-level atoms are always in their adiabatic ground state, with any
spontaneous light scattering ignored. Standing waves pulse $i$ with
the time-dependent light shift~\cite{opticalpumpingoperator}
potential $V_i = \Omega _i (t)\cos ({\bf{Q}}_{\bf{i}} \cdot {\bf{r}}
+ \varphi _i )$ acts as a phase grating to atoms. Here ${\bf{Q}}_i =
{\bf{k}}_i^a - {\bf{k}}_i^b $ is the k-vector of the standing wave.
${\bf{k}}_i^a $, ${\bf{k}}_i^b $ are the k-vectors of the traveling
wave light fields, whose electric field amplitudes are referred to
as  $E_i^a$ and $E_i^b$. And $\varphi _i $ determines the position
of the standing wave nodes in the lab frame. We assume the light
fields have Gaussian beam profiles and the atomic sample is located
at the center of each beam. The cross-sections of all the beams are
characterized by a waist size $w_0$, that is much larger than the
size of the atomic sample. We also assume all the ${\bf{k}}_i^a $
and ${\bf{k}}_i^b $ to have significant differences in propagation
directions such that any linear combinations of ${\bf{Q}}_i $s have
magnitude either zero or on the order of $\frac{{2\pi }}{\lambda }$,
with $\lambda$ the optical wavelength.}

\item[A2.] {
Standing wave pulses are in the Raman-Nath regime, e.g., the
positions of atoms are considered frozen during each pulse duration.
We thus can use Dirac delta functions to effectively describe the
temporal profile of the pulses, e.g., $\Omega _i (t) = \theta _i
\delta (t - T_i )$ that is characterized by its strength $\theta _i
$ and the time of arrival $T_i $. $\theta _i $ is referred as the
pulse area. }

\item[A3.]{
The atomic ensemble is assumed to be dilute such that atom-atom
interactions are ignored. The ensemble is assumed to be dilute in
phase space as well so that the single atom picture is applied
without conflicting with quantum statistics. We can thus use a Dirac
ket $|\psi\rangle$ to label the matter-wave state, that obeys the
single atom Schr{\"o}dinger's equation,

\begin{equation}
i\partial _t |\psi\rangle  = (\frac{{{\bf{\hat p}}^2 }}{2} +
\sum\limits_i {\Omega _i (t)\cos ({\bf{Q}}_i  \cdot {\bf{\hat r}} +
\varphi _i )} )|\psi\rangle .\label{waveequ}
\end{equation}

Given a fixed amount of atoms participating in the interaction, the
spatial density distribution of the atomic ensemble is given by
$\rho({\bf{r}},t) = N_{atom} \rho _{{\bf{r}},{\bf{r}}} (t)$, where
$\rho _{{\bf{r}},{\bf{r}}'} (t) =  \langle \psi ^* ({\bf{r}},t)\psi
({\bf{r'}},t)\rangle $ is the single atom density matrix for the
atomic ensemble in real space. Impacts of $\varphi _i$ on the
matter-wave dynamics require only relatively simple algebra, so we
may drop $\varphi _i$ during the derivation and recover their values
if necessary. }

\item[A4.]{
A grating echo refers to the Bragg scattering of probe light pulse
(the last graph in Fig.~\ref{gratingechoTLIfigure}) off the revived
atomic density grating. The amplitude of the grating echo is
evaluated under first order Born approximation. We assume the
evolution of the atomic density grating is not perturbed by the
scattered light during the probing interval. Using a slowly varying
amplitude approximation, the coupling between $E_N^b $ and the probe
light field $E_N^a $ around time $T_N $ is governed by the equation:
\begin{equation}
(\partial _t  - c\partial _{x_{\bf{Q}} } )E_N^b  = i\alpha \omega
N_{atom} \rho _{{\bf{r}},{\bf{r}}} (t)e^{i{\bf{Q}}_N  \cdot
{\bf{r}}} E_N^a. \label{lightprop}
\end{equation}
Here $x_{\bf{Q}} $ measures the distance along the direction
${\bf{Q}}_N $, and $\alpha $ is the dipole polarizability of ground
state atoms at standing wave frequency $\omega$. If we ignore the
light propagation retardation, the overlap between the
Bragg-scattered field with the $E_N^b $ can be written as:
\begin{equation}
E_N^b (T_N )  =  - i\eta \frac{{N_{atom} \alpha }}{{w_0 ^2 \lambda
}}E_N^a (T_N )\int {d^3 {\bf{r}}} \rho ({\bf{r}},T_N)e^{i{\bf{Q}}_N
\cdot {\bf{r}}}, \label{backamp}
\end{equation}
where $\eta $ is a geometric factor of order unity depending on the
overlap between the Gaussian mode $E_N^a $ and $E_N^b $. $E_N^b (T_N
)$ in Eq.~(\ref{backamp}) gives the magnitude of the grating echo,
and is considered to be the output of the TLI.}
\end{enumerate}
%%%%%%%%%%%%%%%%%%%%%%%%%%%%%%%%%%%%%%%%%%%%%%%%%%%%%%%%%%%%%%%%%
\subsection{The free space displacement diagram}
\subsubsection{The displacement operator and the Weyl function
\label{sectionrule}}

We define $g = \eta \frac{{N_{atom} \alpha }}{{w_0 ^2 \lambda
}}E_N^a (T_N )$. Recalling Eq.~(\ref{weyldefine}), we define the
displacement operator $\hat D({\bf{q}},{\bf{x}}) = e^{i({\bf{q}}
\cdot {\bf{\hat r}} - {\bf{x}} \cdot {\bf{\hat p}})}$ and let
$W({\bf{q}}, {\bf{x}},t)={\bf Tr}(\hat \rho(t) \hat D({\bf{q}},
{\bf{x}}))$, so that Eq.~(\ref{backamp}) can be rewritten in a
compact form using the Weyl function:
\begin{equation}
E_N^b (T_N )  =  - ig W({\bf{Q}}_N ,0, T_N). \label{backamp2}
\end{equation}

To calculate the TLI output $E_N^b (T_N )$ , we might integrate the
matter-wave equation Eq.~(\ref{waveequ}) to express $\hat \rho
{{(T}}_{{N}} )$ in terms of $\hat \rho {{(T}}_{{1}} )$. Here instead
we shall take the Heisenberg picture and integrate the equation of
motion for the displacement operator $\hat D({\bf{Q}},{\bf{x}})$
(Appendix A) so that $W({\bf{Q}}_N ,0,T_N)$ can be expressed in
terms of $W({\bf{q}},{\bf{x}},T_1)$. During free-propagations we
have,

\begin{equation}
W({\bf{q}},{\bf{x}},t)   =  W({\bf{q}},{\bf{x}} - {\bf{q}}t,0).
\label{wignerrule1}
\end{equation}

With the notation $T_i^{ + / - } = T_i \pm 0^ +  $ to remove
ambiguities due to delta functions, transformation of
$W({\bf{q}},{\bf{x}},t)$ due to a standing wave pulse is written as:

\begin{equation}
W({\bf{q}},{\bf{x}},{T_i^ +  })  = \sum\limits_n {J_n (2\theta _i
\sin \frac{{{\bf{Q}}_i  \cdot {\bf{x}}}}{2}) W({\bf{q}} -
n{\bf{Q}}_i ,{\bf{x}},{T_i^ -  })}.  \label{wignerrule2}
\end{equation}

\subsubsection{Basic displacement diagrams}
\begin{figure}
\centering
\includegraphics [width=5 in,angle=0] {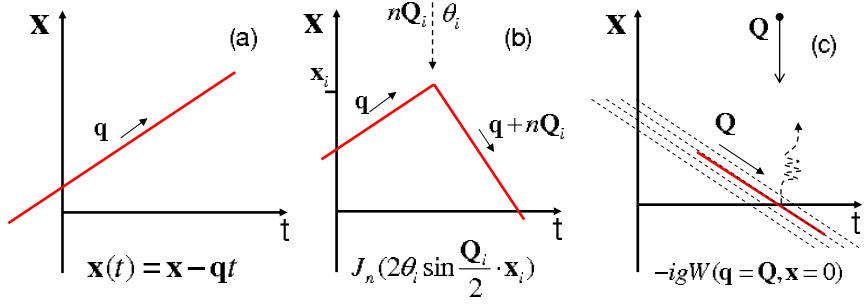}\\
\caption{Basic displacement diagrams: a) A propagating displacement
line [Eq.~(\ref{wignerrule1})]. b) A scattering vertex with an
incoming and outgoing displacement line, the scattering amplitude is
evaluated at the vertex position [Eq.~(\ref{wignerrule2})]. The
dashed arrow represents the action of a standing wave diffraction.
c) A probing vertex that stops an input displacement line
[Eq.~(\ref{backamp2})]. The echo amplitude can be evaluated at the
crossing between the incoming displacement line and the x-axis. The
solid arrow represents the Bragg-scattering (grating echo) process.
The direction of the dashed line array indicates the slope of the
displacement line ${\bf x}(t)={\bf Q}t$.}\label{proprulefig}
\end{figure}

We shall consider Eqs.~(\ref{wignerrule1}) and~(\ref{wignerrule2})
in a space $({\bf q,x})$ that may be called the reciprocal phase
space. From Eq.~(\ref{wignerrule1}), free propagation of
$W({\bf{q}},{\bf{x}})$ is simply a shearing in reciprocal phase
space along $\bf x$. From Eq.~(\ref{wignerrule2}), a standing wave
pulse ``scatters'' $W({\bf{q}},{\bf{x}} ,t)$ into multiple
duplicates shifted along the ${\bf{q}}$ coordinate by $n{\bf{Q}}$
weighted by $J_n (2\theta \sin \frac{{{\bf{Q}} \cdot
{\bf{x}}}}{2})$. We define $ W({\bf{q}}_n (t),{\bf{x}}_n (t)) = (
W({\bf{q}},{\bf{x}},t))_n$, e.g., the component that is shifted by
$n{\bf{Q}}$ due to standing wave scatterings, and represent its
dynamics using a point $({\bf{q(t)}},{\bf{x(t)}})$ that is moving
and scattered in $({\bf{q}},{\bf{x}})$ space. The trajectory can be
projected onto ${\bf{x}}$ coordinates in a ${\bf{x}} - t$ diagram,
where the slope of ${\bf{x}}(t)$ represents the magnitude of
${\bf{q(t)}}$. Equation~(\ref{wignerrule1}) and~(\ref{wignerrule2})
are diagrammatically summarized in Fig.~\ref{proprulefig}(a) and (b)
as the propagation diagram ({\emph{displacement line}}) and
scattering diagram ({\emph{scattering vertex}}) for the Weyl
Functions. In addition, to calculate $E_N^b(T_N)$ fully
diagrammatically, we introduce the probing diagram ({\emph{probing
vertex}}) in Fig.~\ref{proprulefig}(c) to represent the
Bragg-scattering given by Eq.~(\ref{backamp2}).

\subsubsection{The displacement diagram for an N-pulse TLI}
\begin{figure}
\centering
\includegraphics [width=5 in,angle=0] {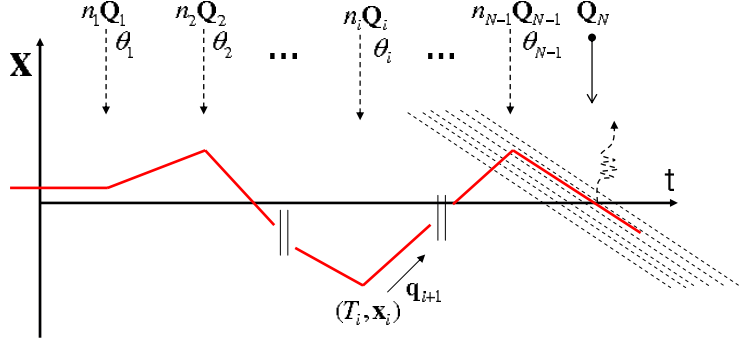}\\
\caption{A diagram corresponding to a particular displacement line
in a general N-pulse TLI.}\label{npulserulefig}
\end{figure}

With further explanations coming shortly, we sketch a representative
path of $({\bf{q}}(t),{\bf{x}}(t))$ that contributes to the grating
echo in the setup sketched in Fig.~\ref{gratingechoTLIfigure}. The
path starts from $({\bf{q}}_1,{\bf{x}}_1)$ at time $t=T_1^-$,
propagate toward $({\bf{q}}_N={\bf{Q}}_N ,{\bf{x}}_N=0)$ at $T_N$,
scattered by N-1 scattering vertex at ${\bf x}_i\equiv {\bf x}(T_i)$
with pulse area $\theta _i $ and standing wave k-vector ${\bf{Q}}_i
$, with diffraction orders $n_i ,i = 1,...,N - 1$ assigned at each
scattering vertex, with $n_N = 1$ and $T_{i,j}  = T_i - T_j $. By
evaluating the Bessel functions at each vertex using the rules in
Fig.~\ref{proprulefig}, the contribution of the path to TLI output
is given by:

\begin{equation}
\begin{array}{l}
E_N^b (T_N ) =   - ig \sum\limits_{\{ n_i \} } {[\prod\limits_{i =
1}^{N - 1} {J_{n_i } (2\theta _i \sin \frac{{{\bf{Q}}_i  \cdot
{\bf{x}}_i }}{2})}]}
W({\bf{q}}_1, {\bf{x}}_1,t={T_1^- }),  \\
with: {\bf{q}}_i  = \sum\limits_{j = i}^N {n_j {\bf{Q}}_j }
,{\bf{x}}_i
= \sum\limits_{j = i}^N {n_j {\bf{Q}}_j } T_{j,i}. \\
 \end{array}\label{npulserule}
\end{equation}
In Eq.~(\ref{npulserule}) we include a summation of all possible
$\{n_i\}$ paths that connect $({\bf{q}}_1 ,{\bf{x}}_1)$ at $t=T_1^-$
with $({\bf{q}}_N={\bf{Q}}_N ,{\bf{x}}_N=0)$ at $T_N$.

Equation~(\ref{npulserule}) can be evaluated with the displacement
diagram in Fig.~\ref{npulserulefig} and the rules given in
Fig.~\ref{proprulefig}. This is the first major result of this work.
To proceed to a calculation based on any initial atomic states, we
still need to specify a particular $({\bf{q}}_1 ,{\bf{x}}_1)$ at
time $T_1^-$.  Clearly, $W({\bf{q}}_1 ,{\bf{x}}_1)$ needs to be
non-zero for a non-vanishing contribution to Eq.~(\ref{npulserule}).
In the next subsection, $W({\bf{q}}_1 ,{\bf{x}}_1 ,t=T_1^-)$ for a
``uniform and broad'' atomic sample is studied and we find that
$({\bf{q}}_1 =0 ,{\bf{x}}_1 \approx 0)$ is required.

\subsection{Momentum conservation and echo condition}
\subsubsection{Input displacement line\label{sectionextensive}}
\begin{figure}
\centering
\includegraphics [width=2in,angle=270] {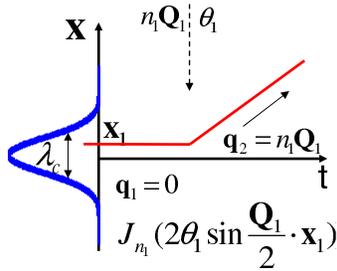}\\
\caption{Diagram restricting the displacement line at the input of
the first scattering vertex [Eq.~(\ref{qconserve}) and
Eq.~(\ref{xconserve1})]. The blue Gaussian on the left represents
the matter-wave spatial correlation function due to Maxwell's
velocity distribution with coherent length $\lambda_c $ . Here ${\bf
q}_1=0$ and ${\mathbf{x}}_1 $ need to be within the coherent length
of the atomic sample.}\label{iniconditionfig}
\end{figure}
This subsection considers the possible values for $({\bf{q}}_1
,{\bf{x}}_1)$ that determine the input displacement line in
Fig.~\ref{npulserulefig}. We shall consider a ``uniform and broad''
atomic sample, specified by the phase-space distribution with the
Wigner function:
\begin{equation}
w_\beta  ({\bf{r}},{\bf{p}}) = g({\bf{r}} - \beta
{\bf{p}})f({\bf{p}}), \label{wigerbeta}
\end{equation}
where $g({\bf{r}})$ is a spatial distribution function, assumed to
be ``uniform and broad'', e.g., uniformly distributed over many
optical wavelengths $\lambda$. This requirement is more general than
simply requiring a distribution over a large spatial extension.
$f({\bf p})$ gives a velocity distribution.

The Weyl function evaluated at $({\bf{q}}_1 ,{\bf{x}}_1)$ is given
by:
\begin{equation}
\begin{array}{c}
W({\bf{q}}_1,  {\bf{x}}_1, t=T_1^-)=< e^{{\bf{q}}_1 \cdot {\bf{\hat
r}} - {\bf{x}}_1  \cdot {\bf{\hat p}}}  > _{T_1^- }\\  = \int {d^3
{\bf{r}}d^3 {\bf{p}}w_\beta ({\bf{r}},{\bf{p}})e^{i({\bf{q}}_1 \cdot
{\bf{r}} - {\bf{x}}_1 \cdot {\bf{p}})} }. \label{wignerbeta2}
\end{array}
\end{equation}

Here ${\bf{q}}_1 $ can only have discrete values specified by
Eq.~(\ref{npulserule}), which is either zero or much larger than the
spatial frequency of $g({\bf{r}})$ according to the assumption (A1)
in section~\ref{sectionmodel1}. A Non-vanishing integration in
Eq.~(\ref{wignerbeta2}) over ${\bf{r}}$ thus requires the condition
of {\emph{momentum conservation}}:
\begin{equation}
{\bf{q}}_1  = \sum\limits_{j = 1}^N {n_j {\bf{Q}}_j }  = 0.
\label{qconserve}
\end{equation}
The restriction from Eq.~(\ref{qconserve}) simplifies
Eq.~(\ref{wignerbeta2}) and we end up with:
\begin{equation}
\begin{array}{c}
W({\bf{q}}_1,  {\bf{x}}_1, t=T_1^-)
= \int {d^3 {\bf{p}}f({\bf{p}})e^{i{\bf{x}}_1  \cdot {\bf{p}}} }\\
={\emph F}({\bf x}_1)\\
= e^{ - 2(\frac{{{\bf{x}}_1 }}{{\lambda_c }})^2 },  \\
 \end{array} \label{wignerbeta3}
\end{equation}
where ${\emph F}({\bf x})$ is the Fourier transform of $f({\bf p})$.
The velocity distribution function $f({\bf p})$ specifies a
coherence length $\lambda_c$. As a simple example, in the third line
of Eq.~(\ref{wignerbeta3}) we consider an isotropic Maxwell
distribution of atom velocity as $f({\bf{v}}) \propto e^{ -
\frac{{{\bf{v}}^2 }}{{2u^2 }}}$, and we define the coherence length
$\lambda_c = \frac{2 \hbar }{{mu}}$. Non-vanishing value of
Eq.~(\ref{wignerbeta3}) requires ${\bf{x}}_1 = \sum\limits_{j = 2}^N
{n_j {\bf{Q}}_j } T_{j,1}$ to be on the order of $\lambda_c $.
Combining with Eq.~(\ref{qconserve}), we have:
\begin{equation}
{\bf{x}}_1  = \sum\limits_{j = 2}^N {n_j {\bf{Q}}_j } T_{j,1} \sim
o(\lambda_c ). \label{xconserve1}
\end{equation}
\begin{figure}
\centering
\includegraphics [width=5 in,angle=0] {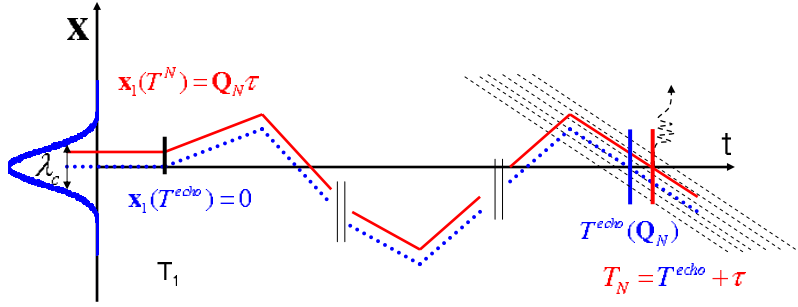}\\
\caption{Echo line in Fig.~\ref{npulserulefig}. Here the black, blue
and red vertical bars indicate the arriving time of the first pulse
$T_1$, the echo time $T^{echo}$, and the arrival time of the
$N^{th}$ pulse $T_N$ respectively.}\label{echofig}
\end{figure}

Equations~(\ref{qconserve}) and~(\ref{xconserve1}) require the input
displacement line in Fig.~\ref{npulserulefig} to start with
$({\bf{q}}_1=0 ,{\bf{x}}_1 \sim o(\lambda_c) )$ in reciprocal phase
space such that $ W({\bf{q}},{\bf{x}}, {T_1^-})$ is non-zero and can
be scattered toward $({\bf{Q}}_N ,0)$ at time ${T_N }$. This is
summarized in Fig.~\ref{iniconditionfig} as a supplementary diagram
to Fig.~\ref{npulserulefig}. With the basic diagrams in
Fig.~\ref{proprulefig} and with the input displacement line
specified in Fig.~\ref{iniconditionfig}, output of an arbitrary
N-pulse TLI can be evaluated using the displacement diagram in
Fig.~\ref{npulserulefig}.

The condition of ${\bf{x}}_1 \sim o(\lambda_c) $ in
Eq.~(\ref{xconserve1}) is different for specific initial condition
with different temperatures, or initial atom velocity spreadings. In
the next subsection, this initial-condition-dependent requirement
will be concretized with an ``echo condition'' for the displacement
diagram.

\subsubsection{Echo condition and echo time}

In the displacement diagram given by Fig.~\ref{npulserulefig}, the
displacement line connects the point $({\bf{q}}_1=0 ,{\bf{x}}_1\sim
0)$ at $t=T_1^-$ with the point $({\bf{q}}_N={\bf{Q}}_N
,{\bf{x}}_N=0)$ at $T_N$. If {$T_i $ , $i=1,...,N-1$} are fixed,
then ${\bf{x}}_1 = {\bf{x}}_1 (T_N )$ is a function of the
detecting-pulse arriving time $T_N$.

With a particular TLI setting described by the parameters $\{ (n_i
{\bf{Q}}_i ,T_i), i = 1,...,N\} $, if ${\bf{x}}_1 $ defined in
Eq.~(\ref{npulserule}) is along the direction of ${\bf{Q}}_N $, then
by varying the time $T_N$, ${\bf{x}}_1 $ can be arbitrarily close to
zero. The {\emph{echo condition}} is defined as follows: In a
general TLI setup [using a ``uniform and broad'' atomic sample
specified by Eq.~(\ref{wigerbeta})] with the first N-1 pulse
parameters set to be $\{ n_i {\bf{Q}}_i ,T_i ,i = 1,...,N - 1\}$, a
revival of atomic density grating with the k-vector ${\bf{Q}}_N $
requires the existence of an echo time $T^{echo} $ such that
\begin{equation}
{\bf{x}}_1 (T_{^N }  = T^{echo} ) = \sum\limits_{j = 2}^N {n_j
{\bf{Q}}_j } T_{j,1}  = 0. \label{xconserve2}
\end{equation}

In the displacement diagram, an echo line is a specific displacement
line that connects $({\bf{q}}_1=0 ,{\bf{x}}_1 = 0)$ and
$({\bf{q}}_N={\bf{Q}}_N ,{\bf{x}}_N=0)$. For the echo line, the
input Weyl function evaluated by Eq.~(\ref{wignerbeta3}) is
maximized. An echo line for the N-pulse diagram in
Fig.~\ref{npulserulefig} is given by the dotted line in Fig
~\ref{echofig}.

\subsubsection{Envelop function of grating echo}

With the echo time defined by Eq.~(\ref{xconserve2}) and with the
momentum conservation condition Eq.~(\ref{qconserve}), it is usually
more convenient to write:
\begin{equation}
{\bf{x}}_1(\tau)   ={\bf Q_N}\tau, \label{equx1}
\end{equation}
with $\tau = T_N - T^{echo}  = \frac{{x_1 (T_N )}}{{Q_N }}$. We can
then express the interferometer output as a function of $\tau $ ,
the time offset of the probing time $T_N$ from the echo time
$T^{echo}$. From Eq.~(\ref{xconserve1}) we see the grating echo
given by Eq.~(\ref{npulserule}) has a $\tau$ dependent envelop
function ${\emph F}(Q_N \tau)$ with its magnitude a symmetric
function with respect to $\tau=0$ and reaches unity at $\tau=0$.

\subsubsection{Total interrogation time}
Notice that when the echo condition Eq.~(\ref{xconserve2}) is
satisfied for $\{T_1=0, T_2, ..., T_N=T^{echo}=T\}$, it must also be
satisfied for $T'\{s_1=0, s_2, ..., s_N=s^{echo}=1\}$, with $s_i
\propto T_i$ for arbitrary $T'$. $T$ shall be called the total
interrogation time of the TLI.

\subsubsection{Vanishing grating echo at the echo time}
In Eq.~(\ref{npulserule}) the only initial condition dependent term
is the last term in Eq.~(\ref{npulserule}). For $T_N=T^{echo}$ such
that $W({\bf{q}}_1 =0 ,{\bf{x}}_1=0)=1$, the TLI output is not only
maximized, but also becomes initial-condition independent. This is
the feature of a ``white light interference'' as the interference is
independent of the degree of input coherence. However, as noticed by
the authors of Ref.~\cite{Periodic02}, if $T_N $ is set to be exact
the echo time $T^{echo}$, the TLI output given by
Eq.~(\ref{npulserule}) vanishes since in the pre-factors of
Eq.~(\ref{npulserule}) $J_{n_1 } (2\theta _1 \sin \frac{{{\bf{Q}}_1
\cdot {\bf{x}}_1 }}{2})$ becomes zero for non-zero $n_1$. Thus TLI
using phase gratings are not ``perfectly-white-light''
interferometers.

The vanishing grating echo at the echo time is understandable, as we
notice the revived fringe at time $T_N$ is a lens-less image of
atomic density grating at time $t=T_N-T^{echo}$ by the following N-2
gratings. If $T_N$ is equal to $T^{echo}$, the atomic distribution
to be imaged is at $t=T_1$, the first pulse arriving time. Since the
first grating is a phase grating, the input matter-waves only
acquire a periodic phase at $t=T_1^+$ but keeps a uniform density
profile. A density grating is developed only after a
$\theta_1$-dependent time during which the atoms bunch to the nodes
of standing wave. The vanishing of the density grating at the echo
time is a feature of TLE using phase gratings. Extension of
interferometry with phase gratings to those with amplitude gratings
will be discussed in the last section of this paper.

For a hot atomic sample with $\lambda_c<<\lambda$, we may Taylor
expand the $J_{n_1}$ term in Eq.~(\ref{npulseruleb}) to have $E^b
\propto \tau^{n_1} {\emph F}(Q_N \tau)$. For $n_1= 1$ and a Maxwell
velocity distribution of $f({\bf p})$, we get a dispersive-shaped
backscattering curve as a function of $\tau$, consistent with
experimental observations~\cite{TLCahn97}.

\subsubsection{Minimum number of pulses in D dimensions}

Finally, we notice that Eqs.~(\ref{qconserve}) and
~(\ref{xconserve2}) are two vector equations. In a particular
experimental setup with fixed grating vectors ${\mathbf{Q}}_i $ and
timing $T_i $ for each standing wave pulses, a non-vanishing grating
echo requires Eq.~(\ref{qconserve}) and Eq.~(\ref{xconserve2}) have
non-zero integer solutions for $\{ n_i ,i = 1,...,N - 1\}$.
Generally, for $\{ {\mathbf{Q}}_i ,i = 1,...,N\} $ which span a $D$
dimension, a minimum number of standing wave pulses are required to
satisfy both equations, given by
\begin{equation}
N_{\min } (D) = D + 2. \label{Nmin}
\end{equation}
For $N > N_{\min }$, more than one set of integer solutions for
$n_i$ may satisfy both equations. The contributions from different
pathways coherently add up and may interfere. An example
illustrating this interference effect (a ``balanced 4-pulse''
configuration) will be given later in this section.

\subsection{Interferometry observable}

\subsubsection{Recoil phase}

We now come to the pre-factor ``$\prod\limits_{i=1}^{N-1}
J_{n_i}(2\theta_i \sin \frac{\bf Q_i \cdot x_i}{2})$'' in
Eq.~(\ref{npulserule}). We shall define:
\begin{equation}
\phi_i^{recoil}=\frac{\bf Q_i \cdot x_i}{2} \label{equrecoilphase}
\end{equation}
as the recoil phase at each scattering vertex in
Fig.~\ref{npulserulefig}, referring to the cumulative phase shift
due to the kinetic energy of recoil effects. Clearly,
$\phi_i^{recoil}$ follows $\bf x_i$ and increases linearly with $T$,
so its slope gives a temporal periodicity of the grating echo
amplitude in Eq.~(\ref{npulserule}). The periodicity of the grating
echo amplitude was measured experimentally to determine the recoil
frequency of atoms~\cite{TLCahn97, hbarBEC02,littleg}.

\subsubsection{Acceleration induced phase\label{accsection}}

We can generalize Eq.~(\ref{npulserule}) in a constantly
accelerating frame (Appendix A). In this case
Eq.~(\ref{wignerrule1}) is modified to include a phase factor:
\begin{equation}
W({\bf{q}},{\bf{x}},t)  =  W({\bf{q}},{\bf{x}} - {\bf{q}}t, 0) e^{i
\frac{m}{\hbar}({\bf{x}} - \frac{{{\bf{q}}t}}{2}) \cdot {\bf{a}}t},
\label{wignerrule1b}
\end{equation}
where ${\bf{a}}$ is the acceleration constant.
Equation~(\ref{npulserule}) is modified accordingly; we have,
\begin{equation}
E_N^b (T_N ) = - ig\sum\limits_{\{ n_i \} } [\prod\limits_{i = 1}^{N - 1} J_{n_i } (2\theta _i \sin \frac{{\bf Q }_i  \cdot {\bf{x}}_i }{2})]  W({\bf{q}}_1 , {\bf{x}}_1 , {T_1^- }) e^{i\frac{m}{\hbar}\int_{T_1 }^{T_N } {{\bf{x}}(t) \cdot {\bf{a}}dt} + \sum\limits_i {n_i \varphi _i } } ,
\label{npulseruleb}
\end{equation}
where ${\bf{x}}(t)$ is given by the displacement line trajectory in
the ${\bf{x}} - t$ diagram as shown in Fig.~\ref{npulserulefig}. In
Eq.~(\ref{npulseruleb}) we also include the phases $\varphi_i$ for
each standing wave pulse explicitly. Equation~(\ref{npulseruleb})
indicates that the displacement line $\bf x(t)$ in
Fig.~\ref{npulserulefig} can be interpreted as the relative
displacement between two diffraction path pair - external
perturbations shift the relative phase between the two paths the
same way as a Mach-Zehnder interferometer.

\subsubsection {Rotation induced phase or Sagnac phase}

Due to the equivalent principle, it is always possible to choose an
inertial frame to describe the atomic motion, and attribute the
inertial forces to the motion of the standing wave field
classically, as long as the space-time curvature is negligible over
the extension of the atomic ensemble. With this in mind, consider
Fig.~\ref{gratingechoTLIfigure} with grating vectors rotated at an
angular velocity $\mathbf{\Omega }$ along a rotation axis across the
atomic sample. In this frame, the standing wave k-vectors $Q_i(T_i)$
is rotated relative to $Q_i(T_1)$ by an amount $\delta
{\mathbf{Q}}_i = {\mathbf{\Omega }} \times {\mathbf{Q}}_i T_{i,1}$.
Accordingly, Eq.~(\ref{qconserve}) is modified to be
\begin{equation}
\begin{array}{c}
\delta {\mathbf{q}}_1 (T_N=T^{echo}) = \sum\limits_{j = 1}^N {n_j
{\mathbf{\Omega }} \times {\mathbf{Q}}_j }
T_{j,1}  \hfill, \\
\delta {\mathbf{x}}_1 (T_N  = T^{echo} ) = \sum\limits_{j = 1}^N
{n_j {\mathbf{\Omega }}
\times {\mathbf{Q}}_j } T_{j,1}^2  \hfill .\\
\end{array} \label{qxconserverotation}
\end{equation}
In Eq.~(\ref{qxconserverotation}) the first line is zero due to
Eq.~(\ref{xconserve2}). We end up with a non-zero displacement
$\mathbf{x}_1$, which can be put together into
Eq.~(\ref{wignerbeta2}) while assuming a mean velocity of the atomic
ensemble to be $< \mathbf{v} >$. We end up with the rotation induced
phase shift for the phase factor in Eq.~(\ref{npulseruleb}):
\begin{equation}
\delta \varphi _{Sagnac}  = ( < {\mathbf{v}} >  \times
{\mathbf{\Omega }}) \cdot \sum\limits_{j = 1}^N {n_j {\mathbf{Q}}_j
} T_{j,1}^2. \label{rotationphase}
\end{equation}

Equation~(\ref{rotationphase}) may also be derived in a rotating
frame and we consider the atomic Coriolis forces. On the other hand,
equation~(\ref{wignerrule1b}) can be derived the same way as
Eq.~(\ref{rotationphase}) by considering the variation of the
standing wave phases due to accelerations.

\subsection {Examples with a single grating vector\label{section1Dexamples}}

\begin{table}[placement]
%\begin{center}  % put inside center environment
\centering
  \begin{tabular*}{1\textwidth}%
     {@{\extracolsep{\fill}}cccr}
TLI name & equal time & balanced  & elongated \\
&   3-pulse & 4-pulse (n) & 4-pulse\\
\hline  % put a line under headers
%%%%%%%%%%%%%%%%%%%%%%%%%%%%%%%%%%%%%%%%%%%%
& & &\\
Ref. Fig. & Fig.~\ref{1Dexpfig1} & Fig.~\ref{1Dexpfig2} & Fig.~\ref{1Dexpfig3} \\
& & &\\
\hline  % put a line under headers
%%%%%%%%%%%%%%%%%%%%%%%%%%%%%%%%%%%%%%%%%%%%
& & & \\
                                                                   & \tiny $J_{-1} (2 \theta _1 \sin \omega_Q \tau )$ & \tiny $ J_{2n - 3} (2\theta _1 \sin \omega_Q \tau )$ &\tiny $J_{-1}(2 \theta _1 \sin \omega_Q \tau)$\\
$ \rho_{\bf Q}e^{2(\frac{v_Q \tau}{\lambda_c})^2-i \varphi}$  & \tiny $J_2 (2\theta _2 \sin \omega_Q(\frac{T}{2}+\tau))$ & \tiny $ J_{4 - 3n}(2\theta _2 \sin (\omega_Q(\frac{(2n - 3)T}{4}+\tau)))$  & \tiny $J_{1}(2\theta _2 \sin \omega_Q (T_s+\tau))$ \\
                                                                   &                                                               & \tiny $J_n (2\theta _3 \sin ( \omega_Q(\frac{T}{4}+\tau)))$ &\tiny $J_{1} (2\theta _3 \sin \omega_Q (T_s+\tau) )$\\
& & &\\
\hline  % put a line under headers
%%%%%%%%%%%%%%%%%%%%%%%%%%%%%%%%%%%%%%%%%%%%
& & & \\
$\varphi(a)/Q a$ & $\frac{1 }{4} T^2$ & $\frac{{6-3n }}{{16}}T^2$  & $T_s (T - T_s )$\\
& & & \\
\hline  % put a line under headers
&&&\\
${\tilde q}/Q\omega_{l}^2$ & $\frac{1}{4}  T^2$ & $\frac{{6-3n}}{{16}} T^2$  & $  T_s(T-T_s)$ \\
&&&\\
\hline  % put a line under headers
  \end{tabular*}
%  \end{center}
\caption{Interferometry setup name (first row), reference figure
(second row), amplitude (third row) and acceleration induced phase
shifts (fourth row) of the $\rho_{\bf Q}$ revivals in
Figs.~\ref{1Dexpfig1},~\ref{1Dexpfig2},~\ref{1Dexpfig3}.
$v_Q=\frac{\hbar Q}{m}$, $\omega_Q=\frac{\hbar Q^2}{2 m}$,
$\lambda_c=\frac{2\hbar}{m u}$ and $u$ is the average thermal
velocity of an atomic sample. The amplitudes of the revival is a
product of all the Bessel functions in each column. In the third
column the ``balanced 4-pulse'' has multiple loop contributions
labeled with index $n$ -- the total amplitude is a coherent
summation over $n$ contributions including the phases. $\tilde q$
(fifth row) gives the modification of revived k-vector due to a weak
quadratic perturbation $V=\frac{1}{2}m \omega_l^2
x^2$.}\label{tableTLIexamples0}
\end{table}

\begin{figure}
\centering
\includegraphics [width=2in,angle=270] {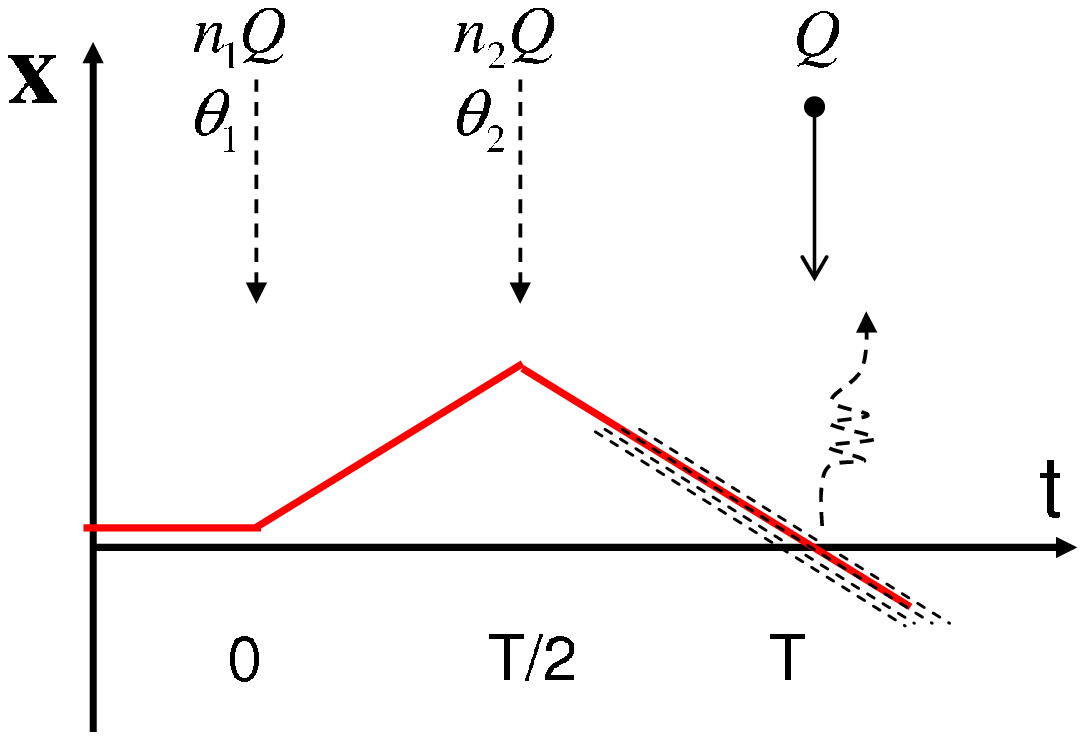}\\
\caption{Diagram corresponding to an ``equal time 3-pulse''
TLI.}\label{1Dexpfig1}
\includegraphics [width=2in,angle=270] {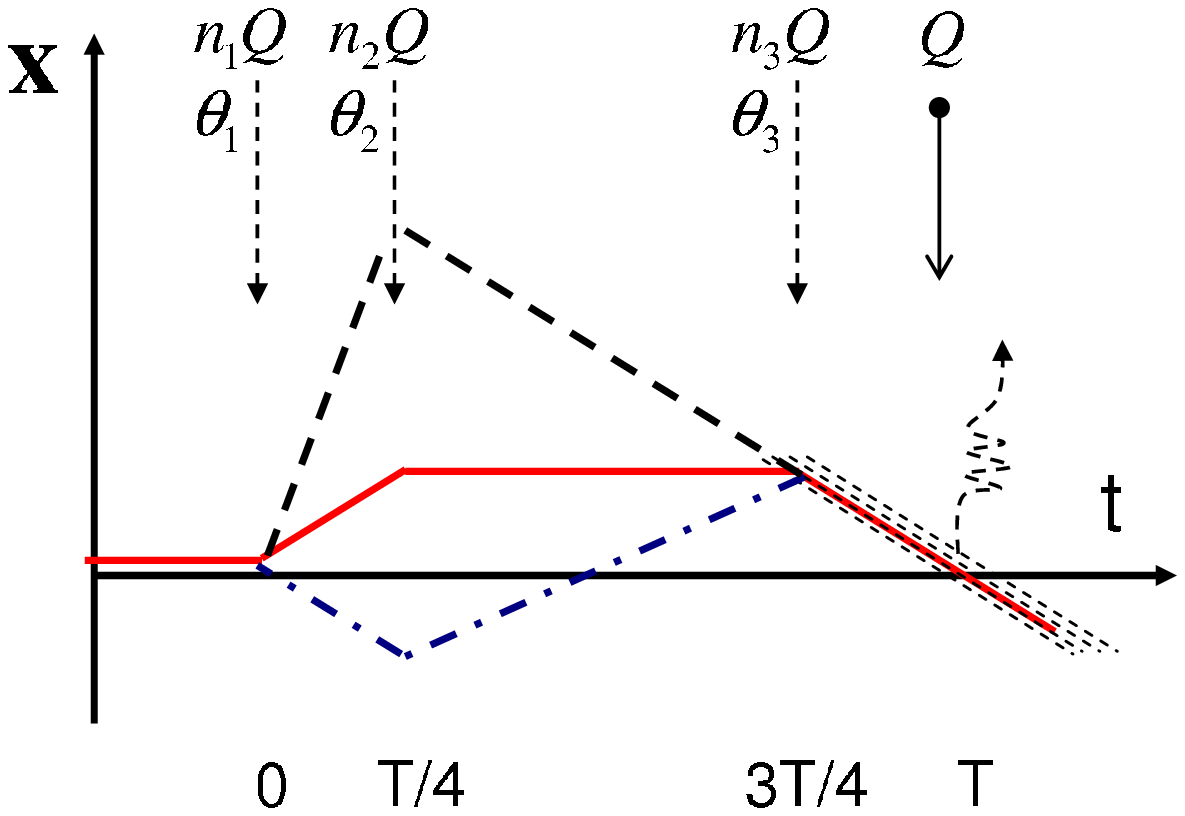}\\
\caption{Diagram corresponding to a ``balanced 4-pulse''
TLI.}\label{1Dexpfig2}
\includegraphics [width=2in,angle=270] {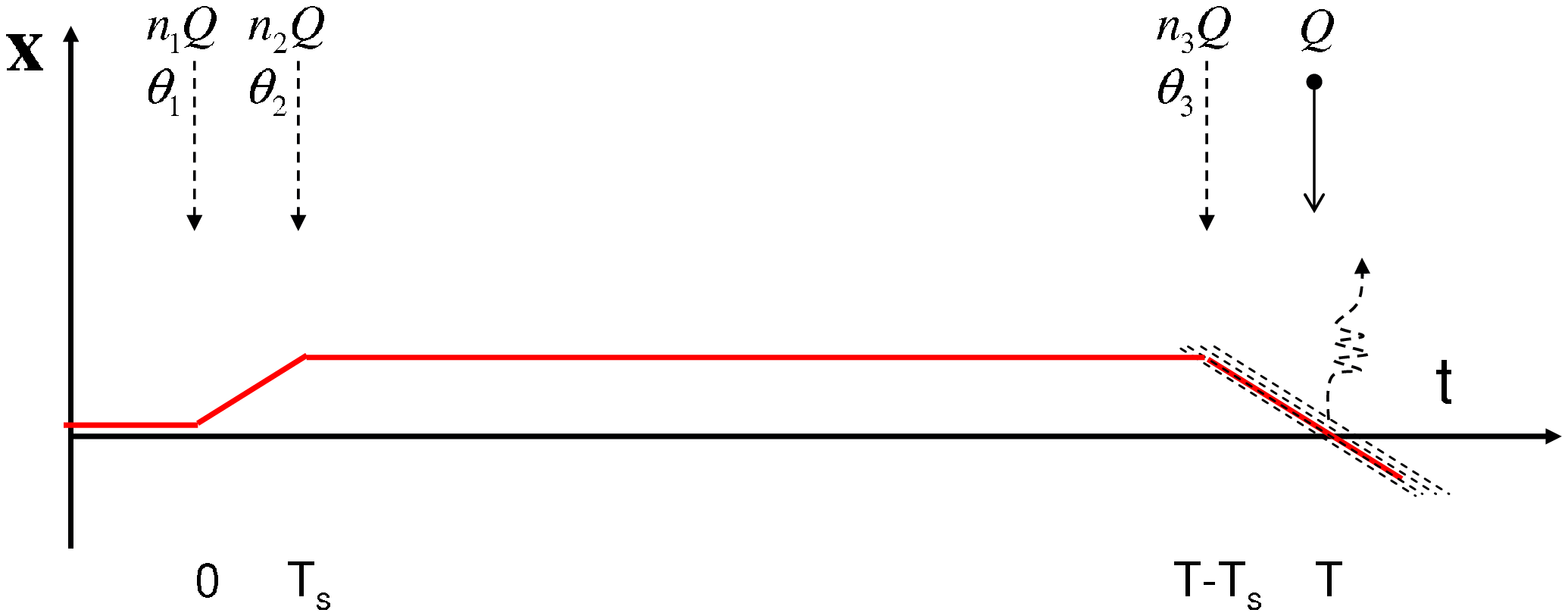}\\
\caption{Diagram corresponding to an ``elongated 4-pulse''
TLI.}\label{1Dexpfig3}
\end{figure}

In this subsection we give three examples of displacement diagrams
for calculating TLI outputs. The most common experimental setup is
with N standing wave pulses from a single standing wave field. In
this case there is only a single grating vector, e.g. ${\bf Q}_i  =
Q {\bf e_x}$. From Eq.~(\ref{Nmin}) we have $N_{\min } (1) = 3$, so
a minimum number of 3 gratings is required to form a grating echo.
In this subsection three $D=1$ TLI schemes will be discussed using
the displacement diagrams~\cite{Periodic02, movingguide,
longcoherence,littleg}.

Figure~\ref{1Dexpfig1} corresponds to a TLI configuration that is
most frequently explored in previous works~\cite{TLCahn97}. In this
configuration, standing wave is pulsed at time $T_1  = 0$, $T_2  =
T/2$, before the revived density grating is probed at time $T_3 = T
+ \tau $. Equations~(\ref{qconserve}) and (\ref{xconserve2}) require
$n_1 = -1,n_2  = 2$. To be specific, we shall call this
configuration an ``equal time 3-pulse'' sequence.

Figure~\ref{1Dexpfig2} corresponds to a ``balanced 4-pulse'' TLI
configuration~\cite{movingguide}, with the standing wave pulsed at
$T_1  = 0$ , $T_2 = T/4$ and  $T_3 = 3T/4$, before the revived
density grating is probed at $T_4  = T + \tau $. Similar timing
sequences have been explored in Ref.~\cite{Periodic02} in three
configurations, namely, the ``fast echo'', ``slow echo'' and the
``stimulated echo''. The ``balanced 4-pulse'' configuration is
understood as a timing sequence such that the ``fast echo'' and the
``stimulated echo'' in ref.~\cite{Periodic02} become degenerate. As
a consequence, more than one paths contributes to the revived
density grating. Equations~(\ref{qconserve}) and (\ref{xconserve2})
require $n_1 = 2n - 3,n_2  = 4 - 3n,n_3 = n$ , where n is an
arbitrary integer. The contribution from different paths add up
coherently in Eq.~(\ref{npulserule}).

Figure~\ref{1Dexpfig3} corresponds to an ``elongated 4-pulse''
TLI~\cite{longcoherence}, with the standing wave pulsed at $T_1 = 0$
, $T_2 = T_s$ and $T_3 = T-T_s$, before the revived density grating
is probed at $T_4  = T + \tau $. The ``elongation'' corresponds to
the timing parameters $T>>T_s$. We shall follow the terminology in
Ref.~\cite{Periodic02} so that the ``elongated 4-pulse''
configuration is considered as an ``elongated stimulated echo''.
Equations~(\ref{qconserve}) and (\ref{xconserve2}) require $n_1 =
-1,n_2  = 1,n_3  = 1$.

The calculation of the interferometry output for the three
configurations in
Figs.~\ref{1Dexpfig1},~\ref{1Dexpfig2},~\ref{1Dexpfig3} is
straightforward using the rules in Fig.~\ref{proprulefig}. The
amplitude and phase of the revived $\rho_{\bf Q}$ are summarized in
Table~\ref{tableTLIexamples0}, where in the last row we also include
a k-vector modification factor due to the perturbation from a
quadratic potential along $\bf e_x$. The latter will be explained in
detail in the next section.

\subsection {A two-grating-vector example\label{section2Dexamples}}
\begin{figure}
\centering
\includegraphics [width=2.2in,angle=270] {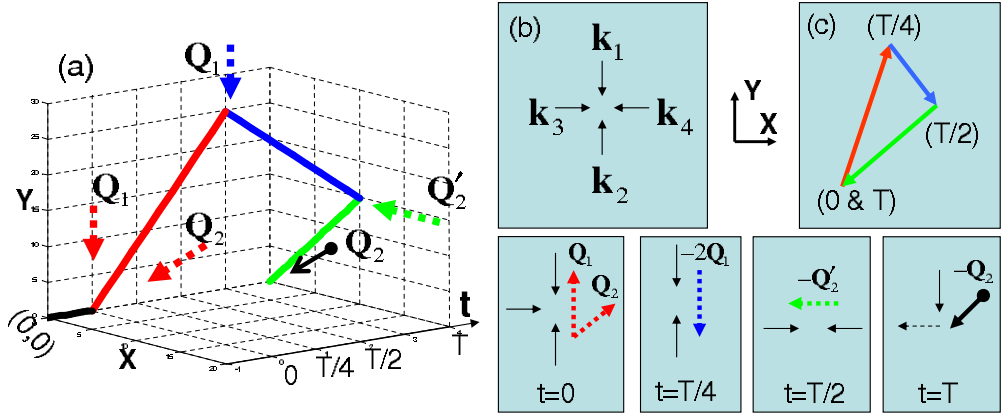}\\
\caption{Diagram corresponding to a 2D TLI. (a) The (X,Y)-t
displacement diagram. (b) The arrangement of the four traveling wave
k-vectors. (c) The projection of the displacement diagram into the
X-Y plane (with arrows to indicate the direction of propagation).
The four graphs on the right bottom gives the relevant standing wave
k-vectors at $t=0$, $T/4$, $T/2$, and $T$. For convenience of
discussion in the main text, the four diagrams are labeled as (d),
(e), (f), (g).}\label{2Dexpfig}
\end{figure}

The examples in the last subsection have been restricted to one
dimension. To complete the discussion on the application of the
displacement diagram, in this subsection we discuss a grating echo
Talbot-Lau interferometry sequence, that involves a manipulation of
atomic coherence using standing wave pulses with the k-vectors span
two dimensions.

As a particular example, we consider a 2D standing wave setup, that
involves two sets of counter-propagating traveling waves with the
k-vectors orthogonal to each other [see Fig.~\ref{2Dexpfig}(b)]. In
particular, we consider the traveling wave light fields labeled with
$E_{1,2,3,4}$, with k-vectors specified as ${\bf k}_{1,2,3,4}$.

According to Eq.~(\ref{Nmin}), a minimum number of N=2+2=4 pulses is
required to complete the TLI sequence in the two dimensional case.
An example sequence is given in Fig.~\ref{2Dexpfig}(a) with a 2D
displacement diagram. The displacement diagram is a 3D plot that
involves two spatial dimensions (reciprocal phase space X-Y) and a
time dimension (t). For clarity purpose, the displacement line after
each vertex is drawn in a different color. The standing wave
k-vectors are colored according to the color of the displacement
line after the scattering vertex. The projection of the displacement
lines into the X-Y plane is shown in
Fig.~\ref{2Dexpfig}(c)~\cite{foot:exp1}. The standing wave pulse
sequence is given in Fig.~\ref{2Dexpfig}(a) as well as in
Fig.~\ref{2Dexpfig} d,e,f,g. In this particular example, the atoms
are diffracted by two standing waves at time $t=0$, with k-vectors
${\bf Q}_1={\bf k}_2-{\bf k}_1=2k \bf e_y$ and  ${\bf Q}_2={\bf
k}_3-{\bf k}_1=k({\bf e_x+e_y})$. The pulse area are $\theta_1$ and
$\theta_1'$ respectively. The second pulse at $t=T/4$ is with a
single grating vector ${\bf Q_1}$ and a pulse area $\theta_2$. The
third pulse at $t=T/2$ is with a single grating vector ${\bf
Q}'_2=2k \bf e_x$ and a pulse area $\theta_3$. The grating echo
induced at time $t=T+\tau$ is due to a probe light from $E_1$ mode
that is Bragg-scattered to $E_4$ mode. Notice that according to
Eq.~(\ref{xconserve2}) we generally have ${\bf x}_1={\bf Q}_N \tau$.
By applying Eq.~(\ref{npulserule}) to Fig.~\ref{2Dexpfig}, it is
straightforward to write down the expected TLI output at time
$T+\tau$:

\begin{equation}
\begin{array}{l}
E_4 (T + \tau ) = i g J_1 (2\theta _1 \sin 2\omega _k \tau ) J_1
(2\theta '_1 \sin 2\omega _k \tau ) \\
J_{ - 2} (2\theta _2 \sin 2\omega _k (3T/8 + \tau )) J_{ - 1}
(2\theta _3 \sin 2\omega _k (T/2 + \tau )) {e^{ - 2({{v_k \tau }
\over {\lambda _c }})^2 } },
\end{array}\label{equ42D}
\end{equation}
with $\omega_k=\frac{\hbar k^2}{2m}$ and $v_k=\frac{\hbar k}{m}$.

\section {TLE in a quadratic potential}

In the last section we have introduced the diagrammatic technique to
calculate a general TLI grating echo that involves free space
matter-waves interacting with N-1 standing wave pulses and a probe
pulse. In this section we extend the technique and consider
matter-waves confined (or anti-confined) in a quadratic potential.

\subsection{The displacement diagram with a quadratic potential}

We shall assume that the  perturbation from the quadratic potential
is ignorable during the standing wave pulse durations. Thus we only
need to modify the (A3) part in the model in
Section~\ref{sectionmodel1}:
\begin{equation}
i\partial _t |\psi \rangle  = (\frac{{{\mathbf{\hat p}}^2 }} {2} +
\frac{1} {2}{\mathbf{\hat r}}^2 :{\mathbf{\omega }}^2  +
\sum\limits_i {\Omega _i (t)\cos ({\mathbf{Q}}_i  \cdot
{\mathbf{\hat r}} + \varphi _i )} )|\psi\rangle
.\label{waveequharmoic}
\end{equation}

For a general discussion, in Eq.~(\ref{waveequharmoic})
${\mathbf{\omega }}$ is a rank two tensor and can be purely
imaginary for an ``anti-trap''.

\subsubsection{Basic displacement diagrams}

The propagation of Weyl functions in a harmonic potential is derived
in Appendix A. We have the modified propagation rule [compare with
Eq.~(\ref{wignerrule1})],

\begin{equation}
W({\mathbf{q}},{\mathbf{x}},t)  =  W({\mathbf{q}} \cdot \cos
{\mathbf{\omega }}t + {\mathbf{x}} \cdot {\mathbf{\omega }} \cdot
\sin {\mathbf{\omega }}t,{\mathbf{x}} \cdot \cos {\mathbf{\omega }}t
- {\mathbf{q}} \cdot {\mathbf{\omega }}^{ - 1} \cdot \sin
{\mathbf{\omega }}t, 0).\label{wignerrule1c}
\end{equation}

\begin{figure}
\centering
\includegraphics [width=2.5in,angle=270] {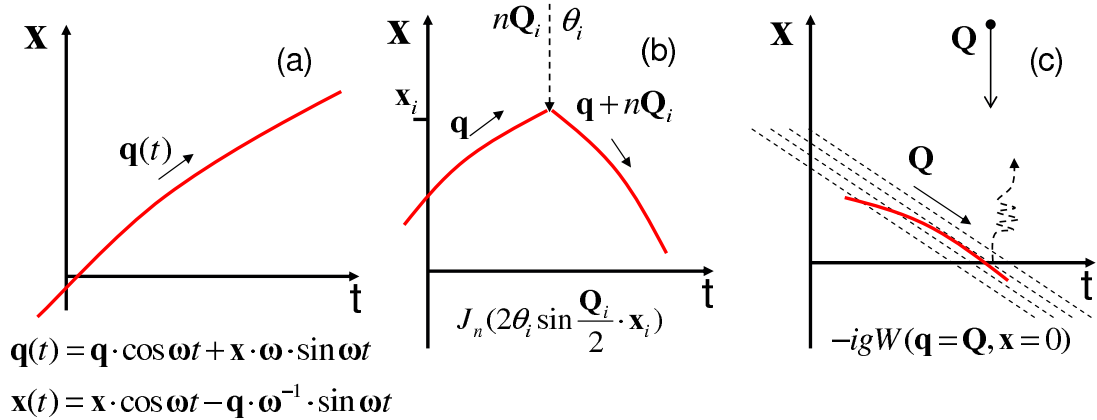}\\
\caption{Modified propagation rules due to Eq.~(\ref{wignerrule1c})
}\label{harmonicfig}
\end{figure}

The modified propagation rule is summarized with the diagram in
Fig.\ref{harmonicfig}(a), together with the scattering diagram and
probing diagram in Fig.\ref{harmonicfig}(b) and (c).
Figure~\ref{harmonicfig} gives the basic displacement diagrams for
TLI in a harmonic trap (compare with Fig.\ref{proprulefig}).

\subsubsection{The displacement diagram for an N-pulse TLI}
\begin{figure}
\centering
\includegraphics [width=5 in,angle=0] {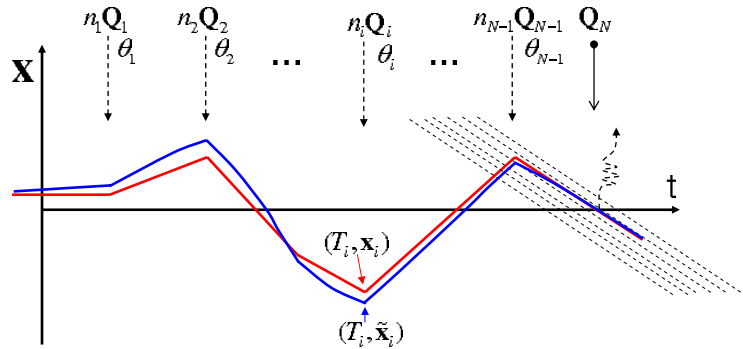}\\
\caption{Diagram giving (${\bf \tilde x}(t)$, ${\bf \tilde q}(t)$)
in a particular path in a general N-pulse TLI under harmonic
perturbation (blue curve). For comparison, the red curve gives a
displacement diagram without harmonic perturbation.
}\label{harmonicnpulsefig}
\end{figure}

In Fig.~\ref{harmonicnpulsefig} we sketch the diagram for a general
N-pulse TLI weakly perturbed by a harmonic potential. The TLI output
can be expressed formally identical to Eq.~(\ref{npulserule}) except
here $\bf{\tilde q}_i$ and $\bf{\tilde x}_i$ need to be evaluated
according to Fig.~\ref{harmonicfig}(a), we have:
\begin{equation}
E_N^b (T_N ) =  - ig\sum\limits_{\{ n_i \} } {[\prod\limits_{i =
1}^{N - 1} {J_{n_i } (2\theta _i \sin \frac{{{\bf{Q}}_i  \cdot
{\bf{\tilde x}}_i }}{2})]}}  W({\bf{\tilde q}}_1, {\bf{\tilde x}}_1,
T_1^-).\label{harmonicnpulse}
\end{equation}

Equation~(\ref{harmonicnpulse}) can be evaluated with the
displacement diagram in Fig.~\ref{harmonicnpulsefig} and the rules
given by Fig.~\ref{harmonicfig}, and is the second major result of
this work. In what follows we discuss TLE due to a harmonic
perturbation in a weak ($\omega T<<1$), pulsed (($\omega (t)\sim
\delta (t)$), and strong ($\omega T>>1$) regime according to
Eq.~(\ref{harmonicnpulse}).

%%%%%%%%%%%%%%%%%%%%%%%%%%%%%%%%%%%%%%%
\subsection{TLE due to weak harmonic perturbations}

\subsubsection{The dephasing factor due to modified input displacement line}
We apply Eq.~(\ref{harmonicnpulse}) to the ``uniform and broad''
atomic sample specified by Eq.~(\ref{wigerbeta}). The Weyl function
in Eq.~(\ref{harmonicnpulse}) is given by
\begin{equation}
\begin{array}{l}
W(\tilde {\bf q}_1, \tilde {\bf x}_1, T_1^-) = \int
g({\bf r}-\beta {\bf p})f({\bf p})e^{i\tilde {\bf q}_1 \cdot {\bf r} -\tilde {\bf x}_1 \cdot {\bf p}} d^3 {\bf r} d^3 {\bf p}\\
={\emph G}(\tilde {\bf q}_1){\emph F}(\tilde {\bf x}_1  - \beta
\tilde {\bf q}_1),
\end{array}
\label{harmonicexten}
\end{equation}
where ${\emph G}({\bf q})$ and ${\emph F}({\bf x})$ are the Fourier
transform of $g({\bf r})$ and $f({\bf p})$ respectively.

The slope of the input displacement line in
Fig.~\ref{harmonicnpulsefig} is modified to be $\tilde {\bf
q}_1(\omega, T)\neq 0$. The grating echo given by
Eq.~(\ref{harmonicnpulse}) has a contrast reduction factor
$C(T)={\emph G}(\tilde {\bf q}_1)$ according to
Eq.~(\ref{harmonicexten}). For example, for a Gaussian or a
Lorentzian distribution of atomic density distribution of a length
$2l$, the contrast reduction factor is given by:
\begin{equation}
\begin{array}{l}
 C(T)_{Gaussian} =   e^{-(\tilde q_1 l)^2/2},\\
 C(T)_{Lorenzian} =   e^{-|\tilde q_1 l|}.
\end{array}\label{equ8dephase2}
\end{equation}

The reduction of the grating echo amplitude as given by
Eq.~(\ref{equ8dephase2}) is due to dephasing, as it stems from the
inhomogeneous phase broadening for matter-wave at different location
in the harmonic potential. Here $\tilde q_1$ is evaluated up to
second order in $\omega T$ for the three 1D examples discussed in
Section~\ref{section1Dexamples} (using Fig.~\ref{harmonicfig}) and
are listed in the last row of Table~\ref{tableTLIexamples0}. An
interesting observation is that $\tilde {\bf q}_1$ has the same $T$
dependence as those in the fourth row that gives the acceleration
induced phase shift. Of particular interest is the $n=2$ path in
Fig.~\ref{1Dexpfig2}. From Table~\ref{tableTLIexamples0} we see
$\tilde q_1 =0$ and the harmonic potential induced dephasing is
greatly suppressed. In this case higher order terms need to be
evaluated, and it's easy to find to the fourth order, $\tilde
q_1=\frac{Q}{64}(\omega T)^4$.

The dephasing described by Eq.~(\ref{equ8dephase2}) is in principle
harmless if we know exactly where and how strong the harmonic
perturbation is, so that we can slightly modify the probing k-vector
${\bf Q}_N$ to compensate for any ``magnification'' or
``demagnification'' of the k-vectors. For a more general potential
that is anharmonic, the dephasing results in a reduction of
interference fringe contrast that cannot be easily compensated. To
evaluate the dephasing factor due to a weak and smooth potential,
one may approximate the anharmonic potential with several local
harmonic potentials.

\subsubsection{Echo profile shift}
\begin{figure}
\centering
\includegraphics [width=2in,angle=270] {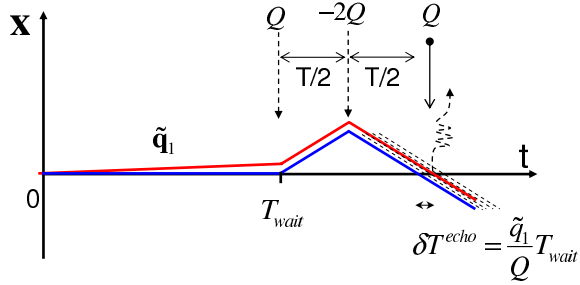}\\
\caption{An example of shifted echo time due to a weak harmonic
confinement. The blue line corresponds to the displacement diagram
in free space, while the red line corresponds to the displacement
diagram due to a harmonic potential.}\label{figechoshift}
\end{figure}

Equation~(\ref{harmonicexten}) also predicts a $\beta$ dependent
shift of echo time. Connected with a typical experimental situation
where a localized thermal atomic sample is expanded during a waiting
time $T_{wait}$, we replace the notation $\beta \rightarrow
T_{wait}$. If $\omega T<<1$ and $T_{wait}=0$ then $\tilde x_1 \simeq
x_1$. For a large $T_{wait}$, the factor ${\emph F}(Q_N \tau -\tilde
q_1 T_{wait})$ in Eq.~(\ref{harmonicexten}) shifts the envelope of
Eq.~(\ref{harmonicnpulse}) as a function of $\tau$ by an amount (we
assume ${\bf Q_N}$ and ${\bf \tilde q}_1$ along the same direction
for simplicity.):
\begin{equation}
\delta T^{echo} = \frac{\tilde q_1}{Q_N} T_{wait}.\label{echoshift0}
\end{equation}

As an example, the echo time shift in the ``equal-time 3-pulse'' TLI
is illustrated with the displacement diagram in
Fig.~\ref{figechoshift}~\cite{foot:exp2}.

\subsubsection{Recoil phase shift}

To finish the discussion of TLE in a weak harmonic potential, we
consider the factor ``$\prod\limits_{i=1}^{N-1} J_{n_i}(2\theta_i
\sin \frac{\bf Q_i \cdot x_i}{2})$'' in Eq.~(\ref{npulserule}). As
discussed in the last section, for a time setting $\{T_1=0,
T_2,...,T\}=T\{s_1=0, s_2, ..., s_N=1\}$ that satisfies the echo
condition Eq.~(\ref{xconserve2}), $E^b$ is a periodic function of
total interrogation time $T$ when there are no harmonic
perturbations.

With harmonic perturbation, ``$\sin \frac{\bf Q_i \cdot x_i}{2}$''
in Eq.~(\ref{npulserule}) is replaced by ``$\sin \frac{\bf Q_i \cdot
\tilde x_i}{2}$'' in Eq.~(\ref{harmonicnpulse}), that in general
breaks the linear relation between the recoil phases in
Eq.~(\ref{equrecoilphase}) and T. However, if the total
interrogation time $T$ is small so that $\omega T<<1$, the linear
relation is still retained~\cite{foot:exp3}. Parallel to
Eq.~(\ref{echoshift0}), we shall consider the atomic sample released
from a local source with a very long $T_{wait}$, so that $\tilde x_i
= x_i +\tilde q_1 T_{wait}$. The amount of recoil phase shift is
given by:
\begin{equation}
\delta \phi_i^{recoil}=\frac{1}{2}{\bf v_Q}_i \cdot {\bf \tilde q}_1
T_{wait},\label{recoilshift0}
\end{equation}
with ${\bf v_Q}_i=\frac{\hbar {\bf Q}_i}{m}$ the recoil velocity.
This shift is systematic in recoil frequency measurements,
particularly if the measurement is carried out in a
trap~\cite{littleg}. The sensitive dependence of the echo shift in
Eq.~(\ref{echoshift0}) and the recoil phase shift in
Eq.~(\ref{recoilshift0}) may also be explored to measure small
curvature of a potential including gravity gradients. Comparing with
the method of using two interferometers separated by a large
distance~\cite{gradiometerTino06,gradiometry02}, monitoring the echo
shift or recoil phase shift of a time domain grating echo
interferometer is more convenient if a field curvature over short
distances is of concern.

\subsection {Pulsed harmonic trap and a time-domain Lau
effect\label{seclaueffect}}

\begin{figure}
\centering
\includegraphics [width=5 in,angle=0] {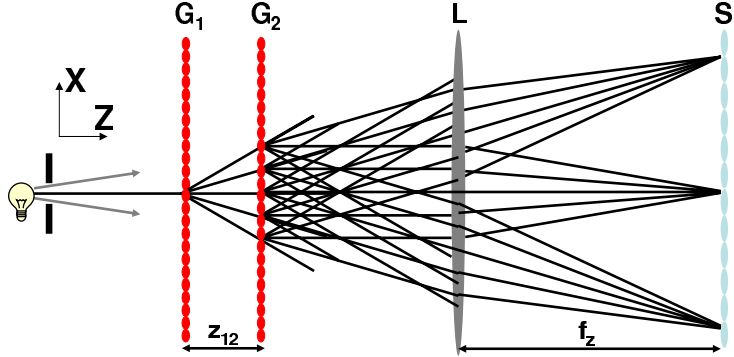}
\caption{A schematic optical layout representing  the setup in Lau's
original discovery~\cite{LauOrigin}. $L$ is an imaging lens at
downstream of two gratings $G_1$ and $G_2$ that are subject to an
incoherent illumination. The screen $S$ is put at the focal plane of
the lens where a periodic image is formed, with its contrast
depending on the separation $a$ between $G_1$ and $G_2$.
}\label{fig6lau}
\end{figure}

In this subsection we shall consider the modified propagation rules
given by Eq.~(\ref{wignerrule1c}) in the short-pulse or the
thin-lens regime. We then apply the result to study the Fraunhofer
diffraction of incoherent illumination over two gratings (the
original discovery of Lau~\cite{LauOrigin}).

We consider a pulse of a one dimensional harmonic trapping potential
with the trap frequency $\omega$ switched on for an interaction time
$\tau_h$, short enough so that $\omega \tau_h <<1$ and the atoms do
not travel significantly in the harmonic trap during the interval
(e.g., the Raman-Nath regime). We may effectively consider
$\omega(t)^2 =f \delta(t)$, with $f=\omega^2\tau_h$.
Equation~(\ref{wignerrule1c}) is reduced to
\begin{equation}
W(q, x,0^+)= W(q - f x, x,0^-). \label{wignerrule1d}
\end{equation}

The pulsed harmonic potential is considered as a thin lens in the
time domain. To see this, consider a paraxial atomic beam with beam
velocity $v_{beam}$. If we make the replacement $z:=v_{beam} t$ and
$f_z:=\frac {v_{beam}}{f}$, the effect of the pulsed harmonic trap
described by Eq.~(\ref{wignerrule1d}) to the atomic beam is
equivalent to a lens with focal length $f_z$.

\begin{figure}
\centering
\includegraphics [width=1.8in,angle=270] {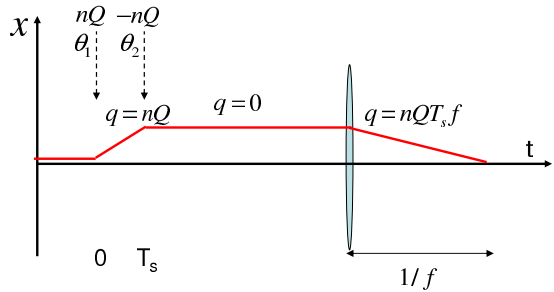}\\
\caption{displacement diagram for a time-domain Lau
effect}\label{Laudiagramfig}
\end{figure}

As a particular application of Eq.~(\ref{wignerrule1d}), in the
following we consider a time-domain version of Lau's original
experiment. We consider an optical layout sketched in
Fig.~\ref{fig6lau}. A un-collimated light passes through two
gratings $G_1$ and $G_2$, with grating constant $d$, separated by a
distance $z_{1,2}$. The Fraunhofer diffraction is observed using a
lens with focal length $f_z$ on a screen `S'. As discussed in the
introduction, a similar setup was used by Ernst Lau in
1948~\cite{LauOrigin, laueffect79} to make the original observation
of the Lau effect: Interference fringe for light at wavelength
$\lambda$ appears on the screen `S' when
$z_{1,2}=n\frac{d^2}{2\lambda}$.

A similar experiment can be implemented in the time domain using
standing wave pulses and a pulsed harmonic potential. To simplify
the discussion, we consider the standing wave with k-vector ${\bf
Q}=Q {\bf e_x}$ and the pulsed trapping potential is also along the
$\bf e_x$ direction. The standing wave is pulsed at time $T_1=0$ and
$T_2=T_s$. The harmonic trapping potential is pulsed at time $T$ at
an amplitude $\omega$ with a short duration $\tau_h$ and we have
$f=\omega^2 \tau_h$. The correspondent displacement diagram is shown
in Fig.~\ref{Laudiagramfig}. We again consider a ``uniform and
broad'' atomic sample [Eq.~(\ref{wigerbeta})] with coherence length
$\lambda_c$. From the diagram we have the revival of an $n^{th}$
order grating:
\begin{equation}
\begin{array}{c}
\rho_{n Q/M}(t=T+1/f+\tau)=W(q=n Q/M,
x=0,t=T+1/f+\tau)=\\J_n(2\theta_1 \sin(\frac{n^2 \omega_Q \tau}{
M}))J_{-n}(2\theta_2 \sin(n^2 \omega_Q (T_s +\tau/M)))e^{-2(\frac{n
v_Q \tau}{M \lambda_c})^2},
\end{array}\label{equlau}
\end{equation}
where $M=\frac{1}{f T_s}$ is a magnification factor that demagnifies
the fringe k-vector from $n Q$ to $n Q/M$ so that the fringe spacing
is magnified by $M$.

We notice the similarity between the ``elongated 4-pulse''
configuration in Fig.~\ref{1Dexpfig3} and the time-domain Lau
configuration in Fig.~\ref{Laudiagramfig}. The time-domain Lau
effect may be most easily realized with a pulsed optical dipole
trap. By reducing $T_s$ and the trap pulse strength $f=\omega^2
\tau_h$, the magnification factor $M$ can be fairly large. In this
way, wavelength-scaled fringes can be converted to a much larger
size, resolvable using regular optical imaging system.

\subsection {TLE in a tight harmonic trap}

\begin{figure}
\centering
\includegraphics [width=2in,angle=270] {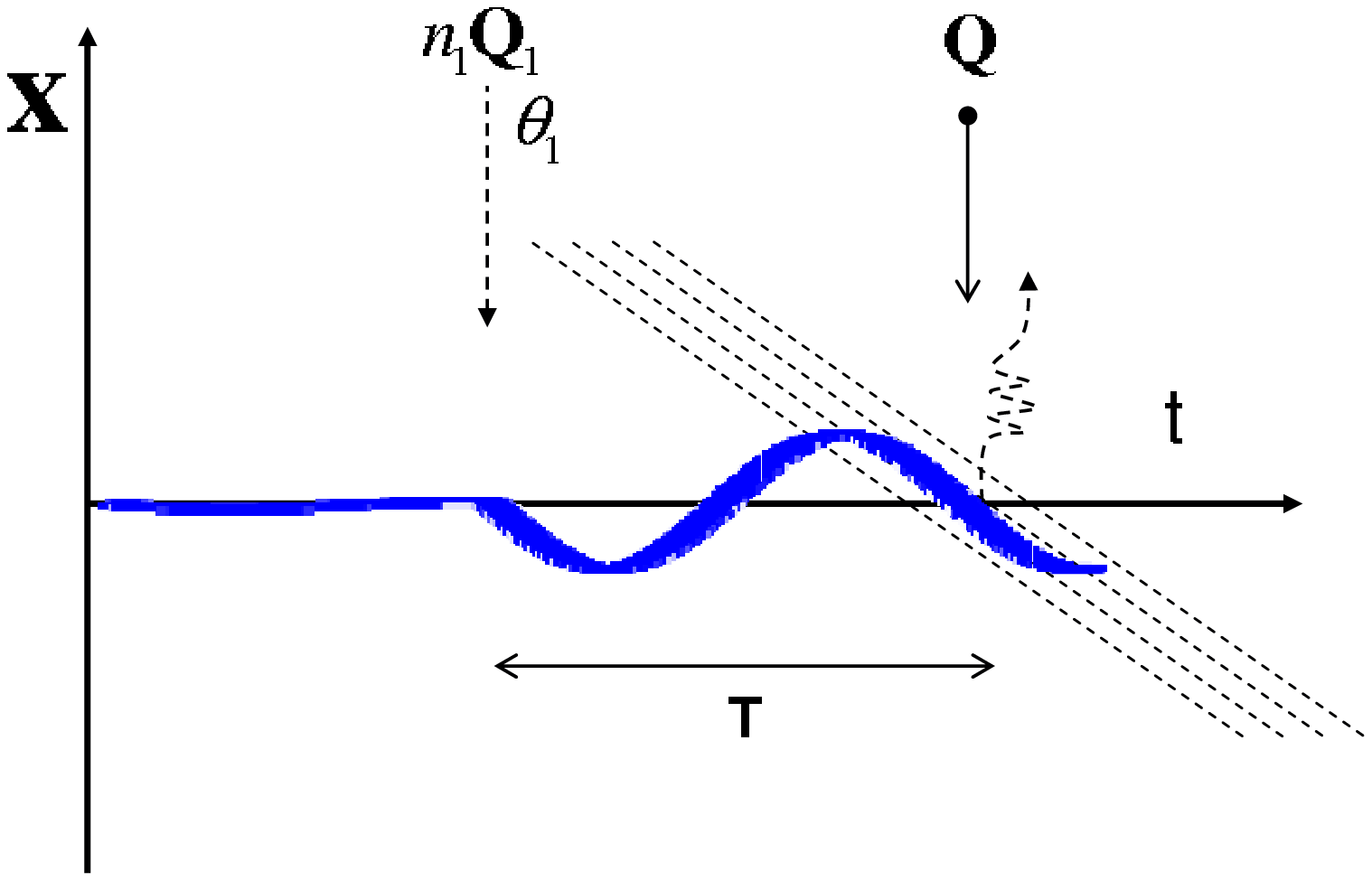}\\
\caption{Diagram of the single pulse TLI in a harmonic
trap}\label{harmonicrevivalfig}
\includegraphics [width=2in,angle=270] {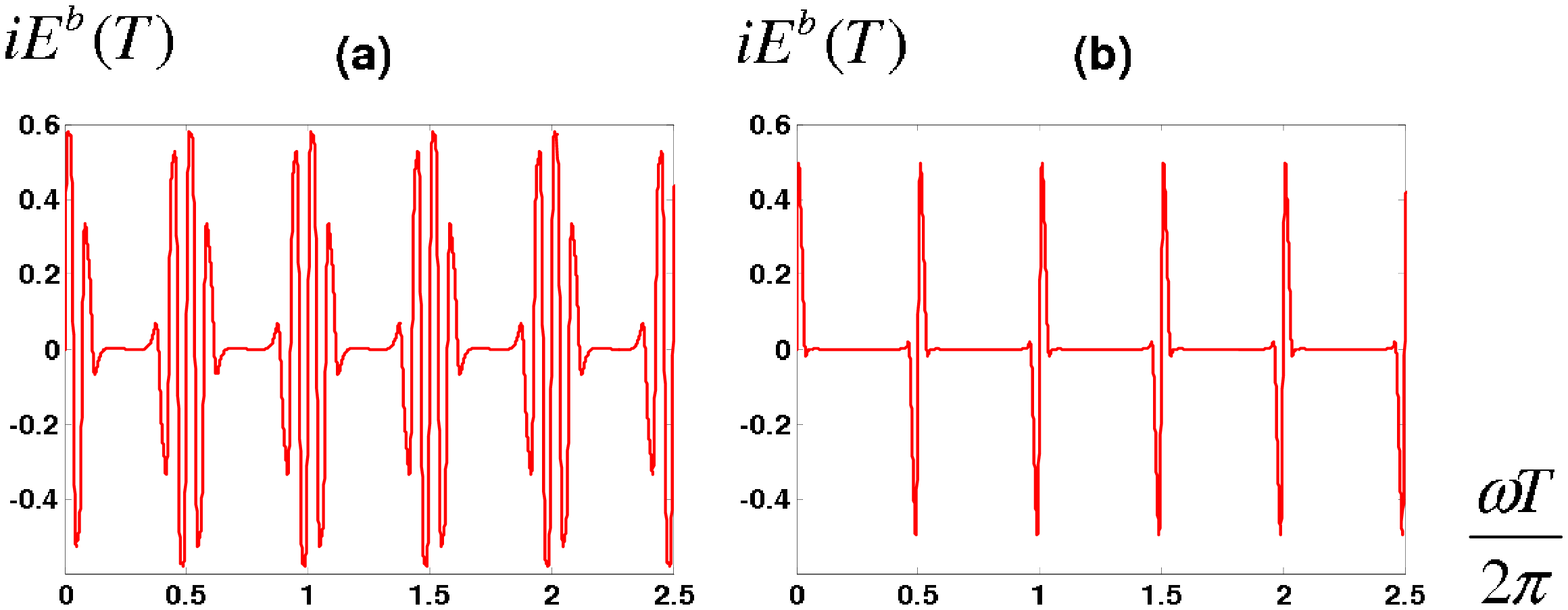}\\
\caption{Representative revived echo signal calculated according to
Eq.~(\ref{harmonicrevival}). The trap frequency $\omega$ is chosen
to be 0.14 times the recoil frequency, $\theta_1=2$. Figure (a)
corresponds to an initial atomic sample at the ground state of the
harmonic potential. Figure (b) corresponds to the initial condition
of a thermal distribution with temperature equal to 50 times the
recoil temperature.}\label{plotrevivalfig}
\end{figure}

In this subsection we consider the revival of atomic density grating
in a harmonic trap. Motion in a harmonic trap is periodic and one
expects the grating echo to revive due to harmonic confinements. In
particular, after a standing wave pulse the atoms bunch toward a
standing wave potential minima to make a density grating, that
repeats itself as the displacement line in
Fig.~\ref{harmonicrevivalfig} crosses the x=0 axis. Consider the
initial atomic distribution to be a Maxwell distribution in the
harmonic trap with spatial extension $L$ and coherence length
$\lambda_c$. We see that in Fig.~\ref{harmonicrevivalfig} $n_1=\pm
1$, and expect a TLI output:
\begin{equation}
E^b (T) =  - ig J_1(2 \theta_1 \sin(\frac{\omega_Q}{\omega}\sin
\omega T)) (e^{-((1-\cos\omega T)Q L)^2}-e^{-((1+\cos \omega T)Q
L)^2}) e^{-(\frac{v_Q}{\omega \lambda_c}\sin(\omega T
))^2},\label{harmonicrevival}
\end{equation}
with $v_Q=\frac{\hbar Q}{m}$, and $\omega_Q=\frac{\hbar Q^2}{2 m}$.

In Fig.~\ref{plotrevivalfig} we plot examples of the atomic grating
revivals. The trap frequency $\omega$ is chosen to be 0.14 times the
recoil frequency $\omega_Q$, while the initial atomic phase-space
distribution is tuned via temperature.
Figure~\ref{plotrevivalfig}(a) corresponds to the atoms initially
occupying the ground state of the harmonic trap only.
Figure~\ref{plotrevivalfig}(b) corresponds to the initial condition
of a thermal distribution with temperature equal to 50 times the
recoil temperature. In both cases the trap frequency determines the
period of the grating revival.

Complimentary to the proposal of using the echo shift
[Eq.~(\ref{echoshift0})] or recoil phase shift
[Eq.~(\ref{recoilshift0})] to measure very weak harmonic
perturbations in the last section, the revival of the atomic density
grating predicted by Eq.~(\ref{harmonicrevival}) may be
experimentally exploited to precisely determine the trapping
frequency of tight traps.

\section {Discussion and Summary}

\subsection{Grating with general transmission functions\label{secgentrans}}

The discussion so far has been restricted to grating potentials
using sinusoidal light shifts. In the following we shall discuss the
generalization of the formula to include arbitrary grating
transmission functions.

We shall first modify the light shift in Eq.~(\ref{waveequ}) to
include an imaginary part, e.g., $\Omega_i+i\Gamma_i=\theta_i \delta
(t-T_i)$, with $\theta_i=\theta_i^R+i \theta_i^I$. The imaginary
part of the light shift potential could be due to the optical
pumping effect of the light pulses~\cite{opticalpumping} that act as
an amplitude mask for atoms. More generally, the imaginary potential
could also describe a particular harmonic component of an absorptive
material grating~\cite{PritchardTL}. Following the principle of
analytical continuation, Equation~(\ref{wignerrule2}) and the
scattering diagram in Fig.~\ref{proprulefig}(b) can be modified to
include this imaginary part as (Appendix A):
\begin{equation}
W({\bf{q}},{\bf{x}},{T_i^+})  =  \sum\limits_n {F (\theta_i,n {\bf
Q}_i,{\bf x})  W({\bf{q}} - n{\bf{Q}}_i ,{\bf{x}},{T_i^-}) },
\label{wignerrule2b}
\end{equation}
where
\begin{equation}
\begin{array}{c}
F(\theta_i,n {\bf Q}_i,{\bf x})=
e^{-2\theta_i^I}\sqrt{\frac{a-b}{a+b}}J_n(\sqrt{a^2-b^2}),\\
a=2 \theta_i^R \sin\frac{{\bf Q}_i \cdot {\bf x}}{2}, b=2 \theta_i^I
\cos\frac{{\bf Q}_i \cdot {\bf x}}{2}.
\end{array}\label{wignerrule2c}
\end{equation}

For a general amplitude/phase grating that has more than one spatial
Fourier component in its transmission function, in
Eq.~(\ref{wignerrule2b}) $F(\theta_i,n\bf {Q_i},\bf {x})$ has to sum
over contribution from all the Fourier components for the $i^{th}$
grating, e.g.,
\begin{equation}
F(\theta_i,n {\bf Q_i},{\bf x})= \sum\limits_{l\times m = n}
{e^{-2\theta_{i}^I(l)}\sqrt{\frac{a_l-b_l}{a_l+b_l}}J_m(\sqrt{a_l^2-b_l^2})},
\label{sumharmonic}
\end{equation}
where $\theta_{i}(l)$ give the effective pulse area due to the
$l^{th}$ order harmonic of the grating potential, $a_l=2
\theta_{i}^R(l) \sin\frac{l{\bf Q_i}\cdot {\bf x}}{2}$ and $b_l=2
\theta_{i}^I(l) \cos\frac{l{\bf Q_i}\cdot {\bf x}}{2}$.

Notice here in
Eqs.~(\ref{wignerrule2b}),~(\ref{wignerrule2c}),~(\ref{sumharmonic})
and also in Eq.~(\ref{wignerrule2}) the subindex $i$ is used to
refer to the $i^{th}$ pulse in a TLI in Fig.~\ref{npulserulefig},
and can in fact be ignored for convenient reading.

To detect the resulting atomic density grating as the output of TLI,
in this paper we have considered the method of a grating echo using
Bragg scattering of a probe light, that is formulated in
Eq.~(\ref{backamp}). Another detection method frequently used for
matter gratings and for resonant standing wave masks is instead to
monitor the overall transmission after an amplitude
grating~\cite{potassiumTL,
PritchardTL,HanschEquivalence05,TLturlapov05, Alexey06}. The formula
developed in this work, e.g., Eq.~(\ref{backamp}) and
Fig.~\ref{proprulefig}(c) can be modified accordingly to
(coherently) sum over the interaction with all the spatial harmonics
of the transmission grating. We end up with the transmission signal
$S(T_N)$ given by [compare with Eq.~(\ref{backamp})]:
\begin{equation}
S (T_N )  \propto \sum\limits_l A_l \int {d^3 {\bf{r}}}  \rho
({\bf{r}},T_N)  e^{i{\bf{l Q}}_N \cdot {\bf{r}}}, \label{sumbackamp}
\end{equation}
where $A_l$ is the coefficient of the $l^{th}$ order spatial
harmonic of the $N^{th}$ grating transmission function.

\subsection{Matter-wave coherence, displacement diagram and recoil diagram\label{secrecoildiagram}}

In this work we discussed using Weyl functions to study the
Talbot-Lau effect (in three dimensions) in free space and in a
quadratic potential. We now come back to Eq.~(\ref{weyldefine}) to
discuss the Weyl function as a measure of matter-wave coherence, and
compare the ``displacement diagram'' technique here with the
``recoil diagram'' introduced by the authors of
ref.~\cite{billiardball82, billiardball93} that has been widely used
in the atom interferometry community~\cite{AIBerman}.

\subsubsection{Displacement diagram and conservation of matter-wave coherence}

We come back to Eq.~(\ref{weyldefine}). As mentioned in the
introduction, $W({\bf q},{\bf x},t)$ measures the overlap, or the
second order field-correlation of matter-waves displaced in phase
space by $(\hbar {\bf q, x})$. As a particular example, $W({\bf
q}=0,{\bf x},t)$ measure the spatial coherence of a wavepacket
before and after a particular displacement $\bf x$, which is also
called the longitudinal coherence in matter-wave
interferometry~\cite{longitudinalcoherence83}.
In~\cite{longitudinalcoherence83} it was pointed out that
longitudinal coherence is conserved during free evolution, which is
a fairly counterintuitive observation since wavepackets of atoms may
expand a lot and one may suspect the coherence length of the
wavepackets increases accordingly.

The conservation of $W({\bf q}=0,{\bf x},t)$ during free evolution
is only a particular example of Eq.~(\ref{wignerrule1}), which more
generally describes the conservation of matter-wave coherence that
propagates in reciprocal phase space $({\bf q,x})$. In particular,
momentum coherence of matter-waves given by $W({\bf q},{\bf
x}=0)=\rho_q$ is a Fourier component of atomic density grating
(fringes), that propagates according to $W({\bf q},{\bf x=-q} t)$
but with conserved magnitude. These propagating coherences reflect
the correlation of matter-waves at different locations with
different velocities, though they cannot be measured directly with
any {\emph{local measurement}}, except those on the axis ${\bf
x}=0$. Equation~(\ref{wignerrule2}) or~(\ref{wignerrule2b}) show how
these phase space coherence can be manipulated with periodic
potentials to create a revived momentum coherence $W({\bf q,x}=0)$
for a local measurement.

Even more generally, it is easy to see from Eq.~(\ref{weyldefine})
the identity~\cite{weylChountasis}
\begin{equation}
\int |W({\bf q}, {\bf x},t )|^2 d {\bf q} d {\bf x}={\bf Tr}(\hat
\rho^2),\label{eqwconserv2}
\end{equation}
which is equal to $<\hat \rho(t)>$, the purity or linear entropy of
matter-waves~\cite{LEdecoherence03, densitymatrixfano}. Matter-wave
entropy is conserved in a non-dissipative/isolated potential, which
may also be interpreted as a conservation of matter-wave coherence
through the Weyl function, given by:
\begin{equation}
\centering \frac{d}{d t}\int |W({\bf q}, {\bf x}, t)|^2 d {\bf q} d
{\bf x}=0.\label{eqwconserv}
\end{equation}

\subsubsection{Displacement diagram and recoil diagram}

\begin{figure}
\centering
\includegraphics [width=2in,angle=270] {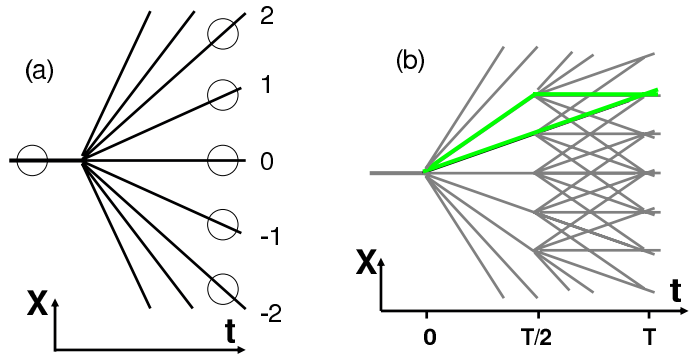}\\
\caption{a) Recoil diagram of matter-waves due to a standing wave
pulse. Circles on each line are ``billiard balls'' with radius equal
to the coherence length of wavepackets (the correspondent
displacement diagram is given by Fig.~\ref{iniconditionfig}). b)
Recoil diagram of a 3-pulse grating echo interferometer (the
correspondent displacement diagram is given by Fig.~\ref{1Dexpfig1})
. Standing wave is pulsed at time $t=0$, $T/2$. At around time $T$
the billiard balls along each line cross each other to create
interference. The circles of ``billiard balls'' are not shown.
}\label{recoildiagramfig}
\end{figure}

The recoil diagram was introduced by the authors of
ref.~\cite{billiardball82} together with a ``Billiard Ball Model''
of atom wavepackets to explain the photon echo phenomena, and also
the Ramsey interference and two-photon echo in matter-wave
interferometry~\cite{billiardball93}. In a (x-t) recoil diagram such
as Fig.~\ref{recoildiagramfig}(a), lines represent the trajectories
of centers of wavepackets diffracted by optical pulses. The
wavepackets are represented with ``billiard balls'' with a fixed
radius equal to the coherence length of the wavepackets. Matter-wave
interferes whenever two billiard balls cross each other. The
visibility of the interfering fringe relies on the overlap of the
billiard balls, while the fringe position is determined by the
relative phase between the two wavepackets. A rigorous justification
of using the ``billiard ball model'' is presented
in~\cite{billiardball82}, which is related to the conclusion in
ref.~\cite{longitudinalcoherence83} and also our previous
discussion.

Examples of recoil diagrams due to a standing wave pulse
(displacement diagram in Fig.~\ref{iniconditionfig}), and for an
``equal time 3-pulse'' (displacement diagram in
Fig.~\ref{1Dexpfig1}) grating echo interferometer is shown in
Fig.~\ref{recoildiagramfig} (a), (b) respectively.  In contrast to
the recoil diagram that concerns the coherence between individual
pairs of wavepackets [such as the green path in
Fig.~\ref{recoildiagramfig}(b)], the displacement diagram describes
the overall coherence of the atomic sample, which is a sum of the
coherence amplitudes between all pairs of the wavepackets with a
same relative phase-space displacement. The displacement diagram is
particularly convenient to describe the overall coherence of
multiple diffraction orders, particularly when a sequence of
multiple diffractions is considered (such as an N-pulse sequence
given by Fig.~\ref{npulserulefig}). The overall coherence of an
atomic sample is directly related to interference readouts in most
interferometry setups where the interference fringes are retrieved
across the whole atomic sample.

\subsubsection{Dephasing and decoherence}

Finally, it is instructive to consider the dephasing effect in both
the recoil diagram picture and in the displacement diagram picture.
For a matter-wave interferometer perturbed by an external potential,
the phase of wavepackets may be in-homogeneously broadened that
reduce the overall coherences of matter-waves and thus the fringe
contrast. This is referred to as a dephasing effect. Dephasing
effects can be described by a recoil diagram, by assigning each
``billiard ball' with a different phase and by varying the ``radius
of the billiard ball'' to account for wavepacket distortions.

In the displacement diagram picture, the dephasing effect broadens
and shifts the Weyl function distribution in the reciprocal phase
space, for example, the dephasing effect due to a quadratic
potential corresponds to a shift of Weyl function described by
Eq.~(\ref{harmonicexten}). However, any broadenings or shifts of
Weyl function distribution need to obey the conservation law given
by Eq.~(\ref{eqwconserv}). In contrast, the reduction of
interferometer fringe visibility accompanied by a violation of
Eq.~(\ref{eqwconserv}) is due to a loss of information or an
increase of entropy in the matter-wave system, and should be
referred to as decoherence effects~\cite{Kothesis, foot:exp4}.

\subsection{Summary and conclusion}

We have generally discussed a vector theory of Talbot-Lau effects
due to N standing wave pulses by considering an N-pulse grating echo
interferometry configuration (Fig.~\ref{gratingechoTLIfigure}). A
displacement diagram (Figs.~\ref{proprulefig} and~\ref{harmonicfig})
is introduced to calculate the grating echo diagrammatically. To
illustrate the convenience of the diagrammatic technique, we discuss
examples of TLI that are related to recent
experiments~\cite{longcoherence,movingguide,littleg}, in particular,
we discuss an example TLI with standing wave k-vector span two
dimensions. We study the dephasing of these TLI schemes due to
harmonic confinements (Table~\ref{tableTLIexamples0}), and predict
the echo shift [Eq.~(\ref{echoshift0})] and the recoil phase shift
[Eq.~(\ref{recoilshift0})] that have been observed
experimentally~\cite{mythesis}. These effects should be useful to
measure the curvature of an atomic potential. The displacement
diagram method is further extended to include a pulsed harmonic trap
[Eq.~(\ref{wignerrule1d})] and we discuss a time-domain version of
the original Lau effect [Eq.~(\ref{equlau})]. Using the displacement
diagram in a tightly confined trap, we give the expression of the
atomic density grating revivals [Eq.~(\ref{harmonicrevival})]. The
formula introduced in this work can be extended to describe TLE due
to diffraction gratings with general grating transmission functions.
Application of Weyl functions in this work has been restricted to
Talbot-Lau effects involving multiple diffraction orders from
diffractive gratings. But the method is generalizable to other
interferometry configurations with fewer interfering paths.

Coherences, or ``off-diagonal terms of a density
matrix''~\cite{densitymatrixfano}, have a clear definition in
discrete systems but not so clear for matter-waves that have a
continuous spectrum~\cite{aireview99}. In the last part of this work
we have briefly discussed Weyl function as a convenient measure of
matter-wave coherence. Weyl function as a phase-space correlation
function measures the overlap of matter-waves before and after a
phase-space displacement. Similar to the definition of the Wigner
function, Equation~(\ref{weyldefine}) involves {\emph {a Fourier
transform of off-diagonal part of a density matrix}}. The Fourier
transform removes the ambiguity due to the ill-defined off-diagonal
terms, while keeping all the physical information of matter-wave
coherences~\cite{foot:exp5}.  We thus suggest using Weyl function to
define matter-wave coherence and more generally to define coherences
in system with a continuous spectrum. We suggest that according to
this definition, a conservation law of matter-wave coherence
[Eq.~(\ref{eqwconserv})] that directly link to the matte-wave
entropy can unambiguously distinguishes decoherence and dephasing
effects. Practically, it seems that to consider the Weyl function or
the displacement operator, instead of considering the evolution of
particular matter-wave states, is more convenient for elucidating
the dynamics of matter-wave coherences during a particular
interferometric observation.

\begin{acknowledgments}
We thank Dr. A. Tonyushkin for careful reading and helpful
suggestions to the manuscript of this paper.
\end{acknowledgments}

\appendix
\section{Propagation properties of the displacement operator}
\renewcommand{\theequation}{A-\arabic{equation}}
\setcounter{equation}{0}

We consider the evolution of the displacement operator $\hat
D(\bf{q},\bf{x})=e^{i(\bf{q} \cdot \bf{\hat r} - \bf{x}\cdot
\bf{\hat p})}$ in the Heisenberg picture. Notice that $\hat
D(\bf{q},\bf{x})$ can also be expressed as $e^{-i \frac{\bf{x}\cdot
\bf{\hat p}}{2}} e^{i{\bf{q}} \cdot {\bf{\hat r}}}e^{-i
\frac{\bf{x}\cdot \bf{\hat p}}{2}}$.

We consider three type of propagators:
\begin{equation}
\begin{array}{c}
U_f(t)=e^{-i (\frac{\bf{\hat p}^2}{2}+\bf {a} \cdot \bf{\hat r})t},\\
U_{harmonic}(t)=e^{-i (\frac{\bf{\hat p}^2}{2}+\frac{\omega^2
\bf{\hat r}^2}{2})t},\\
U_{\bf{Q},\theta}=e^{-i \theta \cos \bf{Q}\cdot \bf{\hat r}}.\\
\end{array}\label{propagators}
\end{equation}

We have, during free propagation in an accelerating frame with
acceleration constant $\bf{a}$,
\begin{equation}
\begin{array}{l}
\hat D({\bf{q}},{\bf{x}},t) = U_f^{ - 1} (t)e^{i({\bf{q}} \cdot
{\bf{\hat r}} - {\bf{x}}\cdot {\bf{\hat p}})} U_f (t) =
\\e^{i \bf{q} \cdot (\bf{\hat r}+\bf{\hat p}t+\frac{\bf{a}t^2}{2}) -
\bf{x}\cdot (\bf{\hat p}+\bf{a}t)}  = \hat D({\bf{q}},{\bf{x}} -
{\bf{q}}t)e^{i (\bf{q}\cdot \frac{\bf{a}t^2}{2}-\bf{x}\cdot
\bf{a}t)}.
\end{array}\label{accderive}
\end{equation}

In a harmonic trap:
\begin{equation}
\begin{array}{l}
\hat D({\mathbf{q}},{\mathbf{x}},t) = U_{harmonic}^{ - 1}
(t)e^{i{\mathbf{q}}
\cdot {\mathbf{\hat r}} + {\mathbf{x}}.{\mathbf{\hat p}}} U_{harmonic} (t) = \\
 e^{i{\mathbf{q}} \cdot ({\mathbf{\hat r}} \cdot \cos
{\mathbf{\omega }}t - {\mathbf{\hat p}} \cdot {\mathbf{\omega }}^{ -
1}  \cdot \sin {\mathbf{\omega }}t) + i{\mathbf{x}} \cdot
({\mathbf{\hat p}} \cdot \cos {\mathbf{\omega }}t +
{\mathbf{\hat r}} \cdot {\mathbf{\omega }} \cdot \sin {\mathbf{\omega }}t)} = \\
 \hat D({\mathbf{q}} \cdot \cos {\mathbf{\omega }}t + {\mathbf{x}}
\cdot {\mathbf{\omega }} \cdot \sin {\mathbf{\omega }}t,{\mathbf{x}}
\cdot
 \cos {\mathbf{\omega }}t - {\mathbf{q}} \cdot {\mathbf{\omega }}^{ - 1}
\cdot \sin {\mathbf{\omega }}t).\\
\end{array}
\end{equation}

Transformation of $\hat D({\bf{q}},{\bf{x}})$ under a standing wave
pulse is given by:
\begin{equation}
\begin{array}{l}
\hat D_+({\bf{q}},{\bf{x}}) = U_{\theta,\bf{Q}}^{ - 1} e^{-i
\frac{\bf{x}\cdot \bf{\hat p}}{2}} e^{i{\bf{q}} \cdot {\bf{\hat
r}}}e^{- \frac{\bf{x}\cdot \bf{\hat p}}{2}}U_{\theta,\bf{Q}}=\\
e^{-i \frac{\bf{x}\cdot \bf{\hat p}}{2}} e^{i{\bf{q}} \cdot
{\bf{\hat r}}}e^{i \theta \cos \bf{Q}\cdot (\bf{\hat
r+\frac{x}{2}})}e^{-i \theta \cos \bf{Q}\cdot (\bf{\hat
r-\frac{x}{2}})}e^{-i \frac{\bf{x}\cdot
\bf{\hat p}}{2}}=\\
e^{- i\frac{\bf{x}\cdot \bf{\hat p}}{2}} e^{i{\bf{q}} \cdot
{\bf{\hat r}}} e^{i 2\theta \sin (\frac{\bf{Q}\cdot \bf{x}}{2}) \sin
(\bf{Q}\cdot \bf{\hat r})} e^{- i\frac{\bf{x}\cdot \bf{\hat
p}}{2}}=\sum\limits_n {J_n(2\theta \sin \frac{\bf Q \cdot
x}{2})}\hat D_-(\bf{q+nQ},\bf{x}).\\
\end{array}\label{scatterderive}
\end{equation}

Finally, we consider the propagator due to a standing wave pulse
with complex area:
\begin{equation}
\begin{array}{c}
U_{\bf{Q},\theta}=e^{-\theta^I+i (\theta^R+i \theta^I) \cos \bf{Q}\cdot \bf{\hat r}},\\
U_{\bf{Q},\theta}^{\dag}=e^{-\theta^I-i (\theta^R-i \theta^I) \cos
\bf{Q}\cdot \bf{\hat r}}.
\end{array}\label{complexprop}
\end{equation}

The transformation of $\hat D({\bf{q}},{\bf{x}})$ is given by:
\begin{equation}
\begin{array}{l}
\hat D_+({\bf{q}},{\bf{x}}) = U_{\theta,\bf{Q}}^{\dag} e^{-
\frac{\bf{x}\cdot \bf{\hat p}}{2}} e^{i{\bf{q}} \cdot {\bf{\hat
r}}}e^{- \frac{\bf{x}\cdot \bf{\hat p}}{2}}U_{\theta,\bf{Q}}=\\
e^{i \frac{\bf{x}\cdot \bf{\hat p}}{2}} e^{i{\bf{q}} \cdot {\bf{\hat
r}}} e^{i 2\theta^R \sin (\frac{\bf{Q}\cdot \bf{x}}{2}) \sin
(\bf{Q}\cdot \bf{\hat r})} e^{- 2\theta^I (1+\cos (\frac{\bf{Q}\cdot
\bf{x}}{2}) cos (\bf{Q}\cdot \bf{\hat r}))} e^{- i\frac{\bf{x}\cdot
\bf{\hat p}}{2}}=\\
e^{-2\theta^I}\sum\limits_n {\sum\limits_m {J_{m+n}(2\theta^R \sin
\frac{\bf Q \cdot x}{2})I_m(2\theta^I \cos \frac{\bf Q \cdot
x}{2})}\hat D_-(\bf{q+nQ},\bf{x})}.
\end{array}\label{complexderive}
\end{equation}
In Eq.~(\ref{complexderive}) $I_n(x)=i^{-n} J_n(i x)$ is the
modified Bessel function of the first kind. From
Eq.~(\ref{complexderive}) to Eq.~(\ref{wignerrule2b}) we use the
Graf summation formula for Bessel functions with complex
arguments~\cite{complexfunc}. The multiple-valued arguments of
Bessel functions in Eq.~(\ref{wignerrule2b}) is chosen to be single
valued according to~\cite{complexfunc}.


\begin{thebibliography}{99}
\bibitem{foot:exp1}
Though a 3D plot as those in Fig.~\ref{2Dexpfig}(a) is necessary for
a 2D displacement diagram, practically we may simply use the
projection plot similar to Fig.~\ref{2Dexpfig}(c).
\bibitem{foot:exp2}
Similar effect was discussed in~\cite{Chugravity} for a
Ramsey-Bord\'e type interferometer under a gravity gradient.
\bibitem{foot:exp3}
A more thorough discussion on the echo profile shift and recoil
phase shift, including an experimental study using interferometry
with magnetically guided atoms, can be found in~\cite{mythesis}.
\bibitem{foot:exp4}
The loss of knowledge to the system may be regained through a
measurement. The process shall still be referred to as a decoherence
effect, as Equation~(\ref{eqwconserv}) has to be violated during
such measurement processes.
\end{thebibliography}
\end{document}